\documentclass[useAMS,usenatbib,a4paper]{mn2e}
\usepackage{amssymb,amstext,amsmath,commath}
\usepackage{times,float,graphicx,paralist}
\usepackage{epsfig,epstopdf,url}
\DeclareGraphicsExtensions{.pdf,.ps,.eps,.png,.jpg,.mps} 
\thinspace
\begin{document}
\title[Star Formation and Environmental Quenching at $z\sim1$] {Star Formation and Environmental Quenching of GEEC2 Group Galaxies at $z\sim1$}
\author[Mok et al.] {Angus Mok$^{1}$, Michael L. Balogh$^{1,2}$, Sean L. McGee$^{2}$, David J. Wilman$^{3}$,
\newauthor Alexis Finoguenov$^{4,9}$, Masayuki Tanaka$^{5}$, Richard G. Bower$^{6}$, 
\newauthor Annie Hou$^{7}$, John~S. Mulchaey$^{8}$, Laura C. Parker$^{7}$\\
$^{1}$Department of Physics and Astronomy, University of Waterloo, Waterloo, Ontario, N2L 3G1, Canada\\
$^{2}$Leiden Observatory, Leiden University, PO Box 9513, 2300 RA Leiden,the Netherlands\\
$^{3}$Max--Planck--Institut f{\" u}r extraterrestrische Physik, Giessenbachstrasse 85748 Garching Germany\\
$^{4} $Department of Physics, University of Helsinki, Gustaf H\"allstr\"omin katu 2a, FI-00014 Finland\\
$^{5}$National Astronomical Observatory of Japan 2-21-1 Osawa, Mitaka, Tokyo 181-8588, JAPAN\\
$^{6}$Department of Physics, University of Durham, Durham, UK, DH1 3LE\\
$^{7}$Department of Physics and Astronomy, McMaster University, Hamilton, Ontario, L8S 4M1 Canada\\
$^{8}$Observatories of the Carnegie Institution, 813 Santa Barbara Street, Pasadena, California, USA\\
$^{9}$CSST, University of Maryland, Baltimore County, 1000 Hilltop Circle, Baltimore, MD 21250, USA\\
}
%
%
\def\etal{{ et al.\thinspace}}
\def\gtrsim{\mathrel{\raise0.35ex\hbox{$\scriptstyle >$}\kern-0.6em\lower0.40ex\hbox{{$\scriptstyle \sim$}}}}
\def\lesssim{\mathrel{\raise0.35ex\hbox{$\scriptstyle <$}\kern-0.6em\lower0.40ex\hbox{{$\scriptstyle \sim$}}}}
\def\Msun{\hbox{$\rm\thinspace M_{\odot}$}}
\def\kmsmpc{{\,\rm km\,s^{-1}Mpc^{-1}}}
\def\kms{km~s$^{-1}$}
\def\oii{[O{\sc ii}]\ }
\def\ewoii{$W_{\rm \circ}$(\oii)}
\def\hdelta{H$\delta\ $}
%
%
\maketitle
\begin{abstract}
\par
We present new analysis from the GEEC2 spectroscopic survey of galaxy groups at $0.8<z<1$. Our previous work revealed an intermediate population between the star-forming and quiescent sequences and a strong environmental dependence in the fraction of quiescent galaxies. Only $\sim5$ per cent of star-forming galaxies in both the group and field sample show a significant enhancement in star formation, which suggests that quenching is the primary process in the transition from the star-forming to the quiescent state. To model the environmental quenching scenario, we have tested the use of different exponential quenching timescales and delays between satellite accretion and the onset of quenching. We find that with no delay, the quenching timescale needs to be long in order to match the observed quiescent fraction, but then this model produces too many intermediate galaxies. Fixing a delay time of 3 Gyr, as suggested from the local universe, produces too few quiescent galaxies. The observed fractions are best matched with a model that includes a delay that is proportional to the dynamical time and a rapid quenching timescale ($\sim0.25$ Gyr), but this model also predicts intermediate galaxies \hdelta strength higher than that observed. Using stellar synthesis models, we have tested other scenarios, such as the rejuvenation of star formation in early-type galaxies and a portion of quenched galaxies possessing residual star formation. If environment quenching plays a role in the GEEC2 sample, then our work suggests that only a fraction of intermediate galaxies may be undergoing this transition and that quenching occurs quite rapidly in satellite galaxies ($\lesssim0.25$ Gyr).
\end{abstract}
%
%
\section{Introduction}
\par
In the local universe, colour-colour diagrams have shown a bi-modality in the distribution of galaxies and a clear minimum between the red and blue peaks \citep{2001AJ....122.1861S}. The bimodal distribution continues to be present at $1\lesssim z\lesssim2.5$, with an increasing contribution from the quiescent galaxies with cosmic time \citep{2008A&A...483L..39C, 2009ApJ...706L.173B}. The well-known `red sequence' is dominated by old, quiescent stellar populations and dust-reddened spirals, while the `blue cloud' is predominately late-type galaxies with ongoing star formation \citep{2004ApJ...608..752B, 2009ApJ...706L.173B}. Finally, the existence of a distinct population of `green valley' galaxies remain controversial \citep{2011MNRAS.412.2303B}.
\par
At $z\sim0$, studies of the stellar mass and luminosity functions of star-forming galaxies have found that they possess a fainter characteristic magnitude and a steeper faint-end slope than passive galaxies \citep{2002MNRAS.333..133M, 2003ApJS..149..289B}. At $z\sim1$, the normalization of the luminosity function for star-forming galaxies has changed little, while it has doubled for quiescent galaxies \citep{2007ApJ...665..265F}. Furthermore, the stellar mass density for star-forming galaxies have remained constant, while it has doubled for passive galaxies since $z\sim1.2$ \citep{2007A&A...476..137A}. All these results suggest that star-forming galaxies must transition to a quiescent state, which then `builds' up the red sequence.
\par
On the whole, star formation peaked at $z\sim2$ and has been decreasing until the present day \citep{2012A&A...539A..31C}. For individual galaxies, this decrease in the star formation rate has been linked to stellar mass, with more massive systems showing sharper decreases in star formation \citep{2005ApJ...619L.135J}. This process is also known as `downsizing', where star formation gradually proceeds from higher mass to lower mass galaxies. On the other hand, galaxies that are actively star-forming continue to show a strong relationship between the specific star formation rate (sSFR) and stellar mass at $z\sim1$, with moderate scatter \citep{2007ApJ...660L..43N, 2012ApJ...754L..14S}. Galaxies that lie above this sequence tend to either host an active galactic nucleus (AGN) or a short-lived burst of star formation, while those below the relation may be cases where the galaxy's star formation is shutting down \citep{2012ApJ...754L..29W}.
\par
Other than epoch and stellar mass, a galaxy's environment can play a significant role in the quenching of star formation. Dense environments tend to have more massive galaxies and their galaxy populations are increasingly dominated by early-type galaxies \citep{2009ARA&A..47..159B}. Observationally, the effectiveness of environmental quenching on star formation has been studied through the work of many photometric \citep[e.g.][]{2012A&A...538A.104G, 2012ApJ...749..150L, 2012ApJ...746...95L, 2013ApJS..206....3S} and spectroscopic surveys \citep[e.g.][]{2009MNRAS.399..715J, 2009A&A...508.1173P, 2011MNRAS.412.2303B}. Even at $z\sim1$, surveys have found that the fractions of galaxies found along the `red sequence' are higher for over-dense regions at fixed stellar masses \citep{2011ApJ...742..125G, 2012ApJ...746..188M, 2012PASJ...64...22T, 2012A&A...539A..55P}. In addition, the existence of radial gradients in the fraction of quiescent and star-forming galaxies inside groups may be further evidence of the work of environmental quenching upon satellite infall \citep[e.g][]{1996Natur.379..613M, 2012A&A...539A..55P}.
\par
Theoretically, many processes have been proposed to explain the effect of environment. Inside groups and clusters, the process of strangulation may occur, where the satellite galaxy is stripped of its hot gas when entering a larger halo \citep{1980ApJ...237..692L, 2000ApJ...540..113B, 2008MNRAS.387...79V}. In very dense environments, ram pressure stripping of the cold gas content of infalling satellite galaxies could become an important factor \citep{1972ApJ...176....1G, 1999MNRAS.308..947A, 2008MNRAS.383..593M}. Finally, gravitational harassment from other group or cluster members may also help to quench star formation in satellite galaxies \citep{2009ApJ...699.1595P}.
\par
One method of interpreting the data is in the context of the halo model, in which the properties of the dark matter halo determines the properties of the galaxies inside. In this model, the evolution of satellite galaxies can be distinguished from that of central galaxies and quenching occurs when satellites enter a halo of a sufficient mass. Therefore, the radial gradients of the quiescent fraction observed in groups and clusters are caused by the satellite galaxies' accretion history into the halo \citep{2000ApJ...540..113B, 2001ApJ...547..609E, 2012MNRAS.419.3167S, 2012MNRAS.423.1277D}.
\par
By trying to match the shape of the observed star formation rate distribution at $z=0$, \citet{2013MNRAS.432..336W} recently proposed that this environmental dependence should take the form of a long delay of t$_Q \sim 2-4$ Gyr, followed by rapid ($\tau_Q<1$ Gyr) satellite quenching timescale. This has also been previously suggested by simulations, where satellite galaxies are often able to retain a significant fraction of their hot gas after infall \citep{2008MNRAS.389.1619F}
\par
Another empirical model of galaxy evolution that seems to account for most of these observations was put forward by \citet{2010ApJ...721..193P, 2012ApJ...757....4P}. In this model, galaxy evolution is driven by two `quenching rates', one factor that depends on stellar mass and another factor that is related to environment, in the form of the local density. While the efficiency of environmental quenching in this model is roughly independent of stellar mass and redshift, its effects should be most prominent amongst low-mass galaxies, $M<10^{10.5}M_\odot$, for which mass-quenching is inefficient.
\par
Therefore, the careful study of environmental dependence remains crucial to our understanding of galaxy evolution. First, the \citet{2010ApJ...721..193P} model and the physical effects of environment quenching both point towards low mass satellite galaxies as an important area of study. Second, groups are an important test of the effects of environment on galaxy evolution, as the majority of galaxies are found inside groups \citep{2004MNRAS.355..769E}. Galaxy groups also provide a crucial link in the hierarchical growth of structures from individual galaxies to dense clusters and observations have shown that galaxies may have been pre-processed in groups before entering the cluster environment \citep{2008ApJ...680.1009W, 2012ApJ...745..106L}. Third, at $z\sim1$, most infalling galaxies would not have spent a long time as a satellite \citep{2009MNRAS.400..937M} and there are high infall rates into groups and the galaxies themselves still possess relatively high star formation rates, which allows us to more easily find any transitional objects that may be moving from the star-forming to the quiescent phase.
\par
This paper will use the data collected as part of the Group Environment Evolution Collaboration 2 (GEEC2) study \citep{2011MNRAS.412.2303B, 2013MNRAS.431.1090M}, a highly complete spectroscopic survey of X-ray selected groups at $0.8<z<1.0$. Using an optical-NIR colour-colour diagram, we found that intermediate (`green') galaxies comprise up to $\sim15-20$ per cent of the group population and possess average star formation rates intermediate between those of the star-forming and quiescent populations. These intermediate galaxies could be a transitional population between the star-forming and quiescent states. Another possible origin for these `green' galaxies is the result of the rejuvenation of early-type galaxies \citep{2012ApJ...761...23F}, which can occur through merger events with smaller gas-rich galaxies. We will attempt to constrain both of these possibilities using a combination of observations and detailed stellar synthesis models.
\par
The structure of the paper is as follows. In \S~2, we present a discussion of the GEEC2 sample with our separation into the quiescent (`red'), intermediate (`green'), and star-forming (`blue') populations. In \S~3, we present the different satellite quenching models, and detail the procedure of using stellar synthesis models to produce intermediate galaxies. In \S~4, the model results are presented, including the predicted quiescent, intermediate, and star-forming fractions as well as their \hdelta strengths. Finally, \S~5 will include a discussion of which models are compatible, the likely constraints based on the observed galaxy properties, and the potential for future observations.
\par
Throughout the paper, a cosmology with $\Omega_m=0.272$, $\Omega_\Lambda=1-\Omega_m$ and $h=H_\circ/\left(100 \mbox{km}/\mbox{s}/\mbox{Mpc}\right)=0.702$ was assumed \citep{2011ApJS..192...18K}. All magnitudes are in the AB system. The stellar synthesis models assume the Chabrier IMF \citep{2003PASP..115..763C}.
%
%
\section{Observations}\label{sec-obs}
\par
Observations from the GEEC2 spectroscopic survey were used. The data-set consists of a spectroscopic survey of $11$ galaxy groups at $0.8<z<1.0$, up to a magnitude limit of $r_{AB}<24.75$. This survey is highly complete ($>66\%$) for $8$ out of the $11$ groups. A detailed discussion of the procedures and results can be found in our previous papers \citep{2011MNRAS.412.2303B, 2013MNRAS.431.1090M}.
\par
The GEEC2 groups in this survey are selected from the X-ray catalog of \citet{2011ApJ...742..125G, 2012ApJ...757....2G}, located in the general COSMOS field \citep{2007ApJS..172....1S}. This dataset is complemented by the 10k zCOSMOS DR2 release, where applicable \citep{2007ApJS..172...70L, 2009ApJS..184..218L}.
\par
The GEEC2 sample has several spectroscopic quality classes, as defined in \citet{2011MNRAS.412.2303B}. Quality class 1 consists of junk spectra and quality class 2 of possible redshifts. Quality class 3 are considered reliable redshifts, including spectra with a good match to the Ca H\&K lines, but no corroborating feature. Quality class 4 is assigned to galaxies with certain redshifts, such as spectra with multiple, robust features. In this analysis, we only use objects with quality class of 3 or above.
\par
Similarly, only the zCOSMOS 10k galaxies with good redshifts are chosen \citep{2007ApJS..172...70L, 2009ApJS..184..218L}. This includes the zCOSMOS quality classes 3, 4, and 9, which corresponds to secure redshifts, very secure redshifts, and one line redshifts (such as the H$\alpha$ or \oii feature). In addition, to maintain consistency, only galaxies located in the same redshift range as the GEEC2 survey ($0.8<z<1.0$) are selected. This sample does not go as deep as the GEEC2 group sample, but the following analysis focuses on bins of galaxies with the same stellar mass.
\begin{table*}
\begin{minipage}{150mm}
	\begin{tabular}{|l|l|l|l|l|l|l|l|l|l|l|}
 	\hline
 	Group & RA & Dec & $z_\mathrm{mean}$ & $N_{mask}$ & $N_{mem}$ &
 $\sigma$ (km/s) & $R_\mathrm{rms}$ (Mpc) & Comp. ($\%$)
 & $M_{\rm dyn}(10^{13}M_\odot)$ \\
		\hline 
40 & 150.414 & 1.848 & 0.9713 & 2 & 15 & $690\pm110$ & $0.34\pm0.04$ & 64 & $11\pm4.7$ \\ 
71 & 150.369 & 1.999 & 0.8277 & 3 & 21 & $360\pm40$ & $0.34\pm0.03$ & 82 & $3.0\pm0.9$ \\ 
120 & 150.505 & 2.225 & 0.8358 & 3 & 31 & $480\pm60$ & $0.78\pm0.07$ & 69 & $12.7\pm4.2$ \\ 
121 & 150.161 & 2.137 & 0.8373 & 3 & 5 & $210\pm60$ & $0.09\pm0.01$ & 100 & $0.2\pm0.2$ \\ 
130 & 150.024 & 2.203 & 0.9374 & 3 & 34 & $600\pm70$ & $0.71\pm0.06$ & 76 & $17.8\pm5.7$ \\ 
134 & 149.65 & 2.209 & 0.9467 & 3 & 23 & $450\pm60$ & $0.97\pm0.07$ & 70 & $13.3\pm4.7$ \\ 
143 & 150.215 & 2.28 & 0.881 & 3 & 20 & $580\pm60$ & $0.23\pm0.03$ & 100 & $5.2\pm1.7$ \\ 
150 & 149.983 & 2.317 & 0.9334 & 4 & 25 & $300\pm40$ & $0.89\pm0.08$ & 75 & $5.5\pm2$ \\ 
161 & 149.953 & 2.342 & 0.944 & 4 & 8 & $170\pm30$ & $0.53\pm0.12$ & 70 & $1.0\pm0.5$ \\ 
213 & 150.41 & 2.512 & 0.879 & 2 & 9 & $260\pm100$ & $0.84\pm0.13$ & 52 & $3.8\pm3.4$ \\ 
213a & 150.428 & 2.505 & 0.9256 & 2 & 8 & $110\pm30$ & $0.62\pm0.09$ & 40 & $0.4\pm0.3$ \\ 
 	\hline
 	\end{tabular}
	\caption{Properties of the eleven galaxy groups observed in the GEEC2 survey, taken from \citet{2011MNRAS.412.2303B}. The position, median redshift $z_{\rm med}$, rest-frame velocity dispersion, and the number of group members are determined from GMOS spectroscopy, combined with available zCOSMOS 10k data. The number of GMOS masks observed in each field is given by $N_{\rm mask}$. $R_{\rm rms}$ is the rms projected distance of all group members from the centre. The spectroscopic completeness of the group is the percentage of candidates within $R_{\rm rms}$ of the group centre for which spectroscopic redshift was obtained, where a candidate is defined as a galaxy with a photometric redshift consistent with the group redshift at the 2$\sigma$ level. Finally, the dynamical mass of the group is listed.}
	\label{tab-gprop}
\end{minipage}
\end{table*}
\subsection{Group and Field Populations}
\par
The group population is taken from the GEEC2 survey and their general properties are shown in Table~\ref{tab-gprop} \citep{2011MNRAS.412.2303B}. The original paper also lists the complete procedure for determining the velocity dispersion, group completeness, distance from the centre, and the dynamical mass of the selected groups, which we will not repeat here.
\begin{table*}
\begin{minipage}{100 mm}
	\begin{tabular}{|l|l|l|l|l|l|l|l|l|}
 	\hline
	& \multicolumn{2}{c}{Star-Forming} & \multicolumn{2}{c}{Intermediate} & \multicolumn{2}{c}{Quiescent}\\
 	Log(Stellar Mass) & Group & Field & Group & Field & Group & Field\\
		\hline 
		9.5 - 9.8 & 23 & 160 & - & - & - & - \\
		9.8 - 10.1 & 20 & 272 & 1 & - & - & - \\
		10.1 - 10.4 & 21 & 211 & 2 & 4 & 3 & 3 \\
		10.4 - 10.7 & 12 & 129 & 7 & 8 & 15 & 15 \\
		10.7 - 11.0 & 13 & 75 & 4 & 19 & 11 & 42 \\
		11.0 - 11.3 & 1 & 23 & 2 & 10 & 11 & 40 \\
		11.3 - 11.6 & - & 1 & - & 5 & 1 & 12 \\
 	\hline
 	\end{tabular}
	\caption{The number of galaxies in each mass bin from the combined zCOSMOS and GEEC2 survey, including the division into star-forming, intermediate, and quiescent.}
	\label{tab-mass}
\end{minipage}
\end{table*}
\par
For field galaxies, our sample is defined as all galaxies in the combined GEEC2 and zCOSMOS 10k catalog not identified as a group member, lying within the range $0.8<z<1$. Note that this field does not represent isolated galaxies, but rather a random selection of representative galaxies. As a result, the field sample should be a fair representation of the `average' population of galaxies at this redshift and provide a good basis for comparison to the group sample. Therefore, the percentage of `group' galaxies would then be the normal percentage expected at this redshift. For example, from the zCOSMOS 10k catalog, this was found to be $\sim20$ per cent for groups with $N\ge2$ and $\sim5$ per cent for groups with $N\ge5$ \citep{2009ApJ...697.1842K}. The exact values depend sensitively on the stellar mass limits of the survey.
\par
A summary of our sample is presented in Table~\ref{tab-mass}, binned by stellar mass, environment, and classification into star-forming, intermediate, and quiescent.
\par
Since this sample mixes together galaxies from two different surveys, a proper comparison will include only data points where both surveys are complete, even for quiescent galaxies with high mass-to-light ratios. For the GEEC2 survey, based on the distribution of mass-to-light ratios, we are complete for quiescent galaxies at $M_{\rm star}>10^{10.5} M_\odot$. For the zCOSMOS survey, the corresponding value for this redshift range is $M_{\rm star}\gtrsim10^{10.7} M_\odot$ \citep{2009ApJS..184..218L}.
\subsection{Galaxy Properties and Classifications}\label{sec-colour}
\par
First, the stellar masses of these observed galaxies are determined by fitting the spectral energy distribution (SED) for each object, using the updated \citet{CB07} models and a \citet{2003PASP..115..763C} initial mass function (IMF). This process is more fully explained in \citet{2011MNRAS.412.2303B}, but the general procedure is to represent the star formation history with a series of exponential decay models with superposed bursts, while varying the other input parameters. The resulting match to the data would not be a single spectrum, but a $\chi^2$ weighted average over the entire series of models. The probability distribution function (PDF) for every parameter of interest is obtained by combining these weights with their values. Then, the median of this probability distribution function (PDF) is taken to be the best estimate of the parameter in question, such as galaxy mass, metallicity, dust attenuation
\par
Second, galaxies are classified into quiescent (`red'), intermediate (`green'), and star-forming (`blue') according to the classification system in our previous paper \citep{2013MNRAS.431.1090M}, where a thorough explanation and justification can be found. The general idea is to separate the quiescent, intermediate, and star-forming galaxies in a colour-colour plane, $(J-[3.6])^{0.9}$ vs $(V-z)^{0.9}$, as shown in Figure 4 of \citet{2013MNRAS.431.1090M}. The superscript indicates that the two colours have been k-corrected to to $z=0.9$, the redshift of interest for our survey, using the {\sc kcorrect} {\sc IDL} software of \citet{2007AJ....133..734B}.
\par
The $(J-[3.6])^{0.9}$ colour is sensitive to dust extinction while the $(V-z)^{0.9}$ colour brackets the 4000 \AA\ break at $0.8<z<1$ and is a good indicator of the luminosity-weighted age \citep{1999ApJ...527...54B}. This particular combination of filters and colours was chosen to avoid contamination from dusty objects into our classification system.
\par
The colour cuts are chosen by first fitting the well-defined star-forming sequence in the colour-colour plane. We then look at the distribution of perpendicular distances of individual galaxy from the blue sequence and find there are three distinct populations, each of which can be well fit with a Gaussian function (see \citet{2013MNRAS.431.1090M}).
\par
Next, we then define the boundaries of the intermediate (`green') population such that it has the minimal overlap with the passive and star-forming sequence. There are additional vertical limits drawn in our colour-colour plot in order to prevent very `blue' or very `red' objects from being identified as intermediate galaxies.
\subsection{Spectroscopic Indices}\label{sec-spec}
\par
Spectral line indices are measured for all galaxies with secure redshifts from the GEEC2 sample. The features measured include the \oii feature, which is used to determine star formation rates and the \hdelta line, which is used as an indicator of recent star formation and one of the primary constraints for our models. The general procedure of measuring equivalent widths and the results from observations of the GEEC2 group galaxies are outlined in our previous paper \citep{2013MNRAS.431.1090M}.
\par
In general, measurements of the equivalent widths are done using the bandpass method. First, the continuum flux levels of the spectra both red-ward and blue-ward of the feature are found. This is done by averaging the flux over the relevant wavelength ranges in the red and blue continuum. Next, the continuum flux level in the area of the spectral line can be defined by drawing a line between the centres of the red and blue continuum ranges. This is done because the continuum levels on the line itself can be hard to measure and define, especially for strong spectral features. The relevant wavelength ranges for the \oii and H$\delta$ features are presented in Table~\ref{tab-cont}.
\par
The contribution from the line feature is determined by integrating the flux over the line index definitions listed in Table~\ref{tab-cont} for each spectral feature and subtracting the continuum estimate. For this paper, we adopt the sign convention such that a positive \oii equivalent width indicates an emission feature, while a positive \hdelta equivalent width corresponds to an absorption feature. Note that we will use the measured strengths of the \hdelta absorption line for each population (after excluding spectra from the telluric contaminated region of $0.818<z<0.895$) from \citet{2013MNRAS.431.1090M}.
\begin{table}
	\begin{tabular}{|l|l|l|l|}
 	\hline
 	Index & Blue Continuum & Line Definition & Red Continuum \\
		\hline 
		\oii & 3653 - 3713 & 3722 - 3733 & 3741 - 3801 \\
		H$\delta$ & 4030 - 4082 & 4082 - 4122 & 4122 - 4170 \\
 	\hline
 	\end{tabular}
	\caption{The table list the wavelength intervals used for the estimation of the blue and red continuum, as well as the line definition for the spectroscopic indices used in this paper. The values in the table are provided in units of Angstroms.}
	\label{tab-cont}
\end{table}
\subsection{Star Formation Rates}\label{sec-sfr}
\par
This section will discuss the star formation rate of the galaxies in the GEEC2 sample and the different measurement methods used in this analysis. It will be a continuation of the discussion on this topic in \citet{2013MNRAS.431.1090M}, focusing on the properties of the group galaxies.
\subsubsection{FUV + IR}
\par
The group galaxies in the GEEC2 sample are located within the larger zCOSMOS survey and many objects have both FUV and MIPS-based Spitzer IR measurements \citep{2007ApJS..172....1S, 2012ApJ...753..121K}. The combination of these two measurements can be one of the best measurements of the total star formation rate of a galaxy, taking into account the UV flux from young stars and the IR emissions from dust attenuation and re-emission.
\par
In order to calculate the FUV + IR SFR, we use the rest-frame k-corrected UV and optical photometric data from the zCOSMOS catalog. To do this, the k-corrections are applied to the filter that most closely matches the rest wavelength of the desired filter, rather than taking the synthetic magnitude from the template fit. The distribution of FUV magnitudes also allows us to obtain a reasonable upper limit for galaxies that are undetected in the rest-frame FUV flux (FUV $\gtrsim 25$ mags).
\par
Next, MIPS 24 micron data is converted into total IR luminosity with the templates from \citet{2001ApJ...556..562C}. The corresponding code can be found at \url{http://david.elbaz3.free.fr/astro_codes/chary_elbaz.html}. Finally, we combine FUV and the total IR luminosities into star formation rate estimates with the prescription of \citet{2011ApJ...741..124H}:
\begin{equation}
L({\rm FUV})_{\rm obs} + 0.46\times L({\rm TIR})_{\rm obs}
\end{equation}
\par
The final conversion between total luminosity and star formation rates is taken from \citet{2007ApJS..173..267S}, for the Chabrier IMF.
\subsubsection{FUV + Dust Attenuation Estimates}
\par
We can also estimate the star formation rates for galaxies which are only FUV detected, but lack the corresponding MIPS 24 micron detection. Since the detection limit for the MIPS data is relatively high, upper limits based on the 24 micron data may be overly conservative. Therefore, we estimate the possible dust attenuation from the UV slope, which would include the use of the NUV band and the (FUV-NUV)$^0$ colour. In line with the convention used in this paper, the superscript indicates that colour has been k-corrected to the rest wavelength.
\par
The relationship from \citet{2009ApJ...700..161S} was used, which had dust attenuation estimated using the (FUV-NUV)$^0$ colour:
\begin{equation}
\begin{split}
A_{\rm FUV}=3.68({\rm FUV-NUV})^0+0.29\ {\rm for}\ ({\rm FUV-NUV})^0<1
\\
A_{\rm FUV}=3.97\ {\rm for}\ ({\rm FUV-NUV})^0\ge1
\end{split}
\end{equation}
\subsubsection{\oii Measurements}
\par
We can also measure the star formation rate of these objects through the \oii feature in their spectra. The \oii luminosity is coupled to the H II regions, which are the sites of star-formation inside galaxies. However, this line is more strongly affected by dust extinction and metallicity than comparable indicators, like the H$\alpha$ recombination feature.
\par
To compensate, \citet{2010MNRAS.405.2594G, 2011MNRAS.412.2111G} presents an empirical conversion between observed \oii luminosities and SFR at $z=0.1$ in a statistical manner, calibrating the result to values derived using H$\alpha$ and UV data:
\par
\begin{equation}\label{eqn-dgg}
\frac{\mathrm{SFR}}{\mathrm{M}_\odot\ yr^{-1}} = \frac{\mathrm{L_{[O\ II]}}}{3.80\times10^{40}\mathrm{erg\ s^{-1}}}\frac{1}{a\ \mathrm{tanh}[(x-b)/c] + d},
\end{equation}
with $a=-1.424,\ b=9.827,\ c=0.572,\ d=1.700,\ x=\log (M_\mathrm{star}/M_\odot)$.
\par
This only includes the average dust correction for galaxies of that stellar mass and does not account for the individual variations in attenuation. It also does not take into account any changes in the properties of these galaxies between $z=0.1$ and $z\sim1$.
\par
To determine the \oii star formation rates for the GEEC2 sample, we calculate the \oii luminosity by multiplying the \oii equivalent width measurement from the spectra with the luminosity of the continuum at its redshifted position, obtained from broad-band photometry. The \oii - based star formation rates had been calculated in \citet{2013MNRAS.431.1090M}, which were then further calibrated using the combination of FUV + IR and FUV + dust attenuation values of the GEEC2 sample.
\par
A comparison of the [OII] dust corrected SFR and the FUV + IR SFR for our sample show a normalization offset, as also noted in \citep[e.g.][]{2011ApJ...735...53P, 2011ApJ...730...61K}. This mass-independent difference has been presented in Appendix 1 of \citet{2013MNRAS.431.1090M} and the \oii - derived star formation rates are multiplied by a factor of 3.1 in this paper as well. Note that calibrating to the entire zCOSMOS field sample produced a smaller normalization value of 2.06. However, this change does not significantly modify our overall result or conclusions. Since the focus of our analysis is on the group population and in order to maintain consistency with past results, we have decided to retain the \citet{2013MNRAS.431.1090M} normalization.
%
%
\begin{figure*}
	\leavevmode
	\epsfysize=5.75cm
	\epsfbox{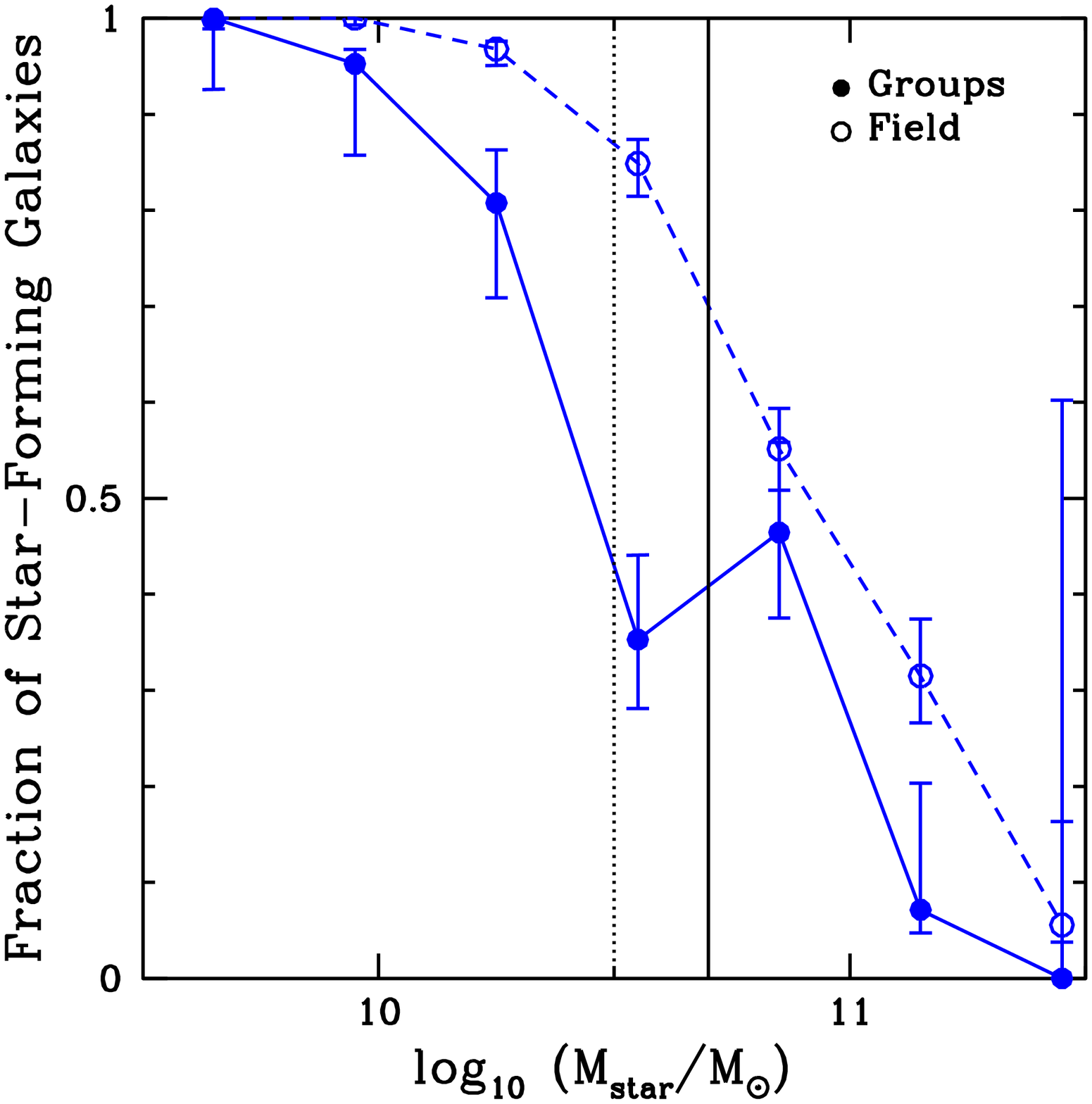}
	\epsfysize=5.75cm
	\epsfbox{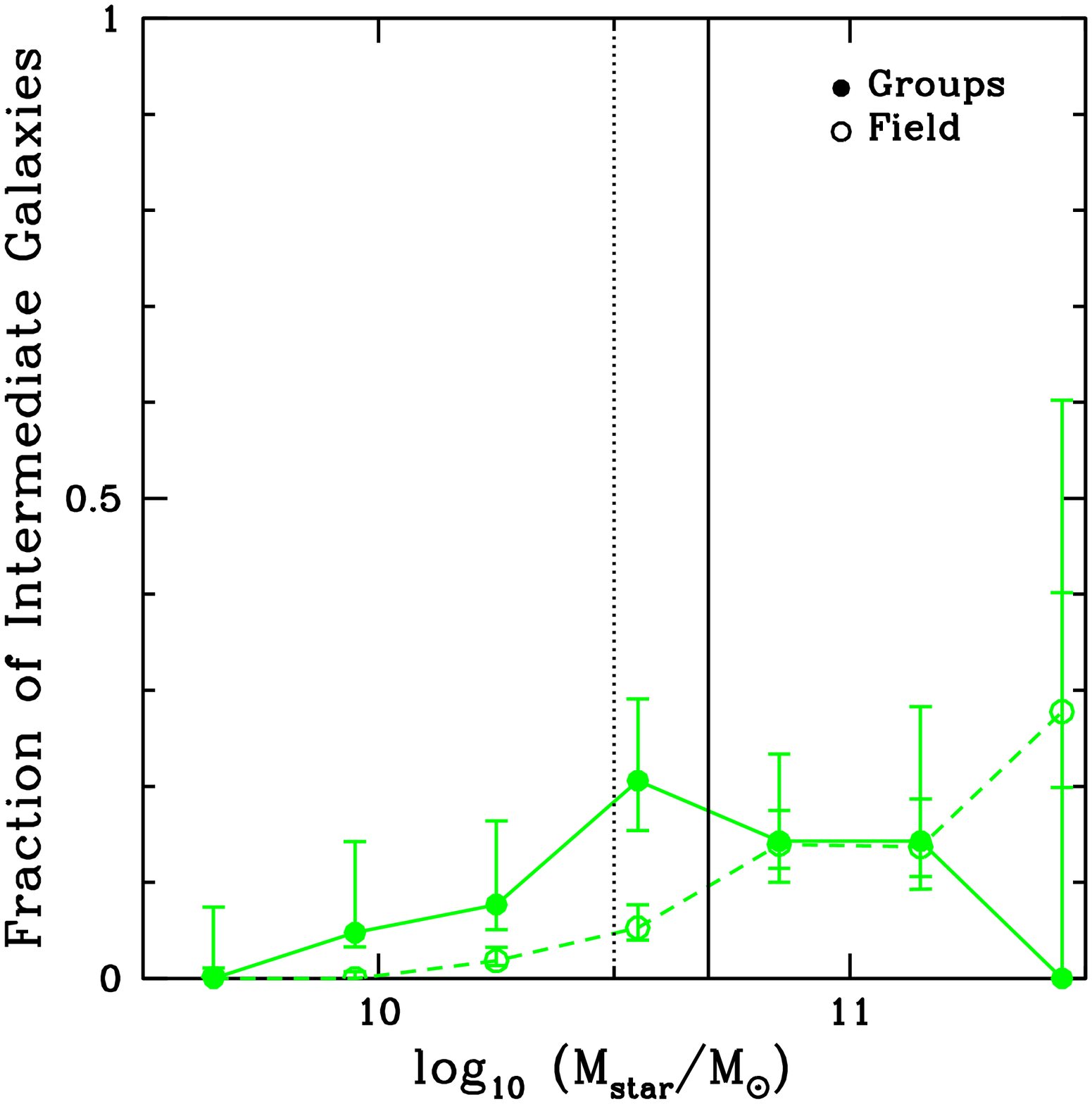}
	\epsfysize=5.75cm
	\epsfbox{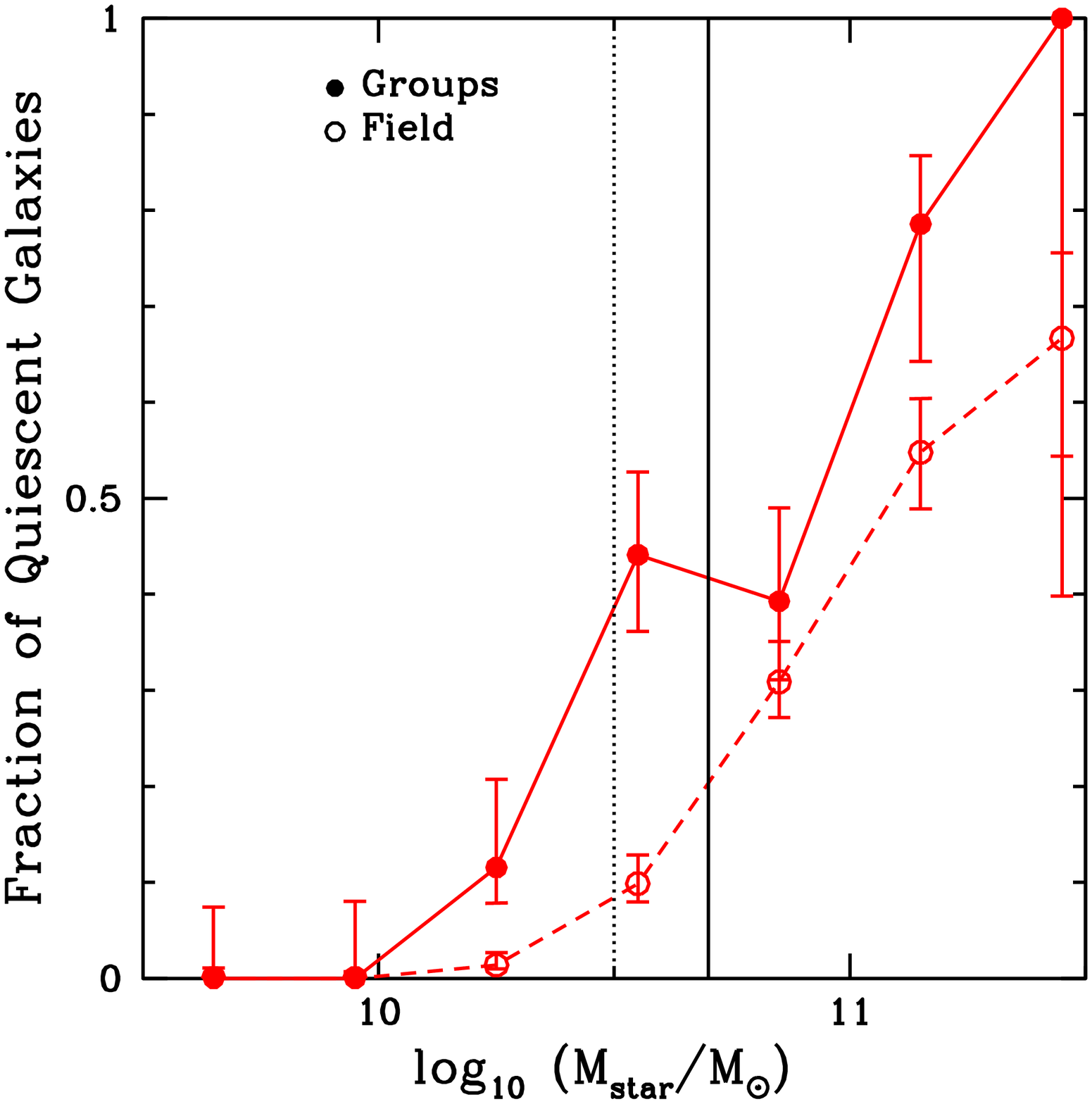}
	\caption{Plots of the fraction of star-forming {\it (left)}, intermediate {\it (centre)}, and quiescent {\it (right)} galaxies in the group and field populations, based on our colour definitions and restricted to galaxies with $0.8<z<1$ and secure redshifts. The group population {\it (filled circles and solid line)} consists of galaxies identified as a group member. The field population {\it (open circles and dashed line)} consists of all galaxies (including objects from the zCOSMOS 10k catalog) not specifically identified as part of a group in Table~\ref{tab-gprop}. Also shown with the data points are their error bars. The black dotted line at $10^{10.5} M_\odot$ shows where the sample is reasonably complete for the GEEC2 sample, even for quiescent (red) galaxies. A similar limit for the zCOSMOS sample is shown with a black solid line at $10^{10.7} M_\odot$.}
	\label{fig-datafrac}
\end{figure*}
\section{Characteristics of Group and Field Galaxies}
\subsection{Star-Forming/Intermediate/Quiescent Fractions}
\par
The fraction of quiescent, intermediate, and star-forming galaxies is shown in Figure~\ref{fig-datafrac}, as a function of stellar mass. The asymmetric error bars are calculated using the binomial error method from \citet{2011PASA...28..128C}. The full analysis can be found in \citet{2011MNRAS.412.2303B} and \citet{2013MNRAS.431.1090M}, but some important results can be noted here.
\par
In \citet{2013MNRAS.431.1090M}, we observed a strong environmental dependence on the fraction of quiescent galaxies, with higher values inside groups than in the field for all stellar masses ranges. In addition, the fraction of intermediate galaxies was found to represent $\sim15-20$ per cent of the overall galaxy population or up to $\sim30$ per cent of the quiescent + intermediate population in the GEEC2 sample. We can also consider the quenching scenario and determine the relative proportion of star-forming and intermediate galaxies. In the region where the two populations are complete, the fraction of intermediate galaxies is $\sim35$ per cent of the population of the combined star-forming + intermediate population. There is also some indication that the fraction of intermediate galaxies is higher inside group, but this is largely driven by stellar masses at which this sample is incomplete.
\subsection{Star Formation Rates}\label{sec-dis_sfr}
\begin{figure*}
	\leavevmode
	\epsfysize=8.5cm
	\epsfbox{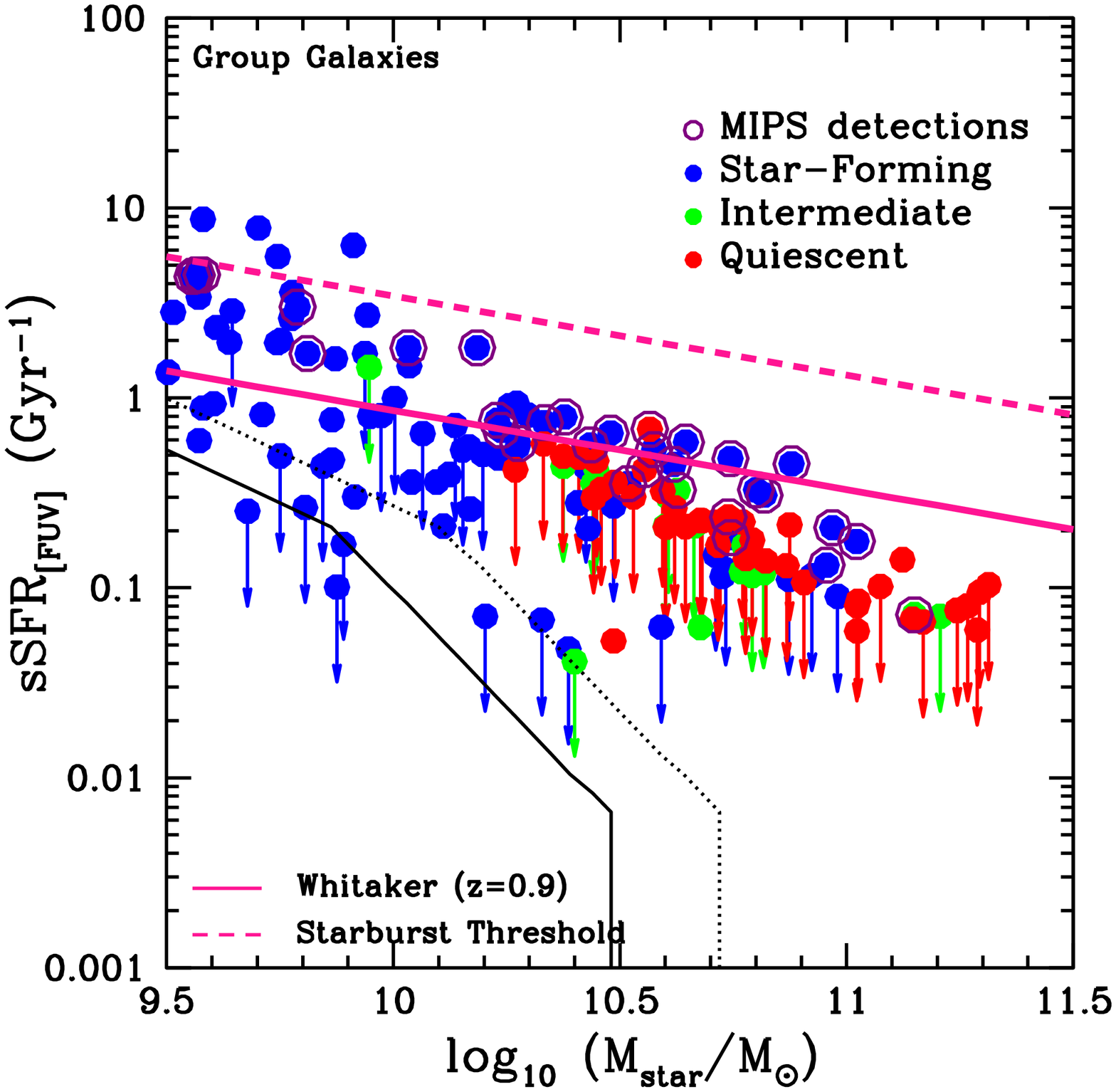}
	\epsfysize=8.5cm
	\epsfbox{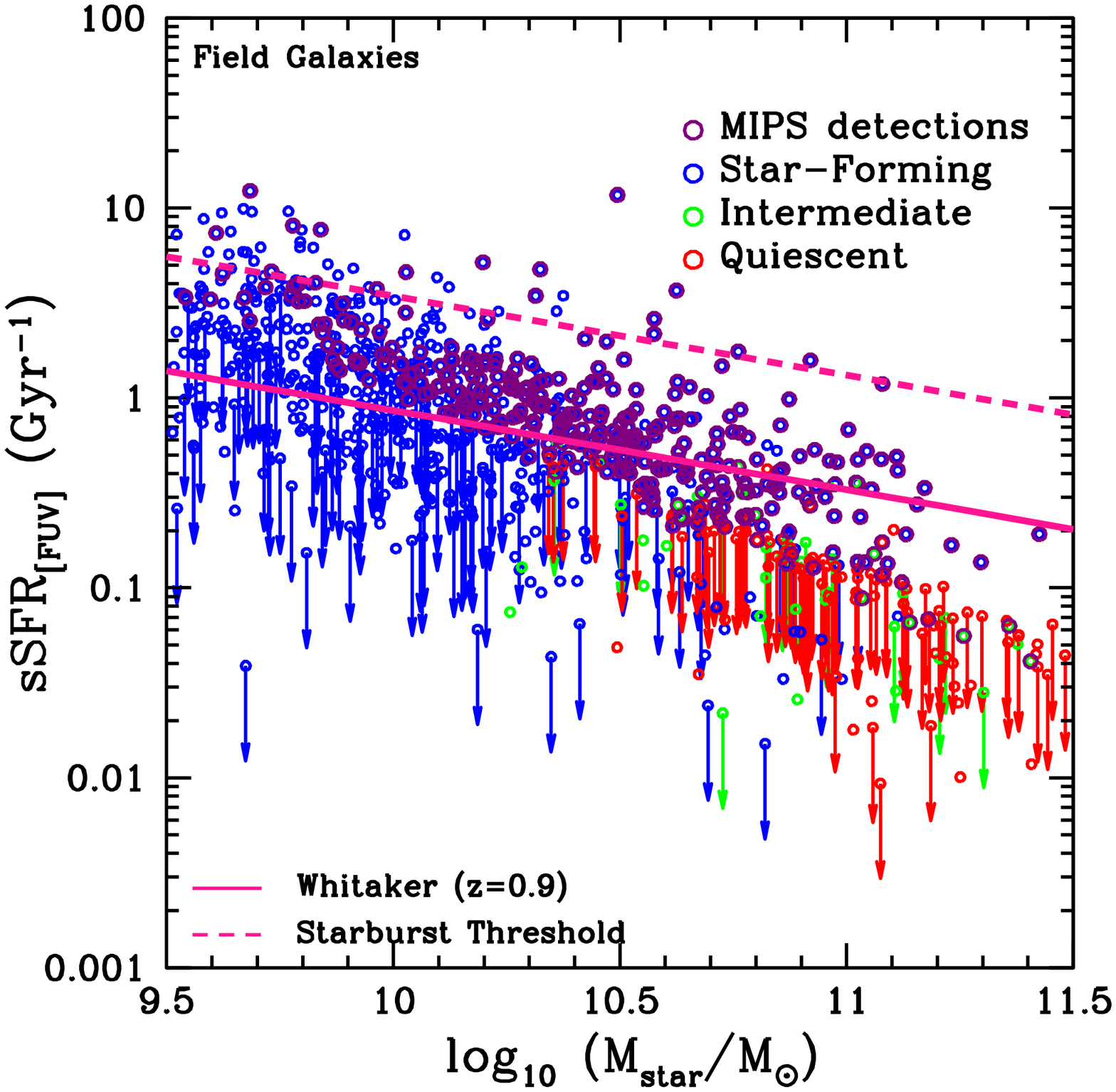}
	\caption{{\it Left:} The specific star formation rate measured from the FUV + IR (where available) and FUV + the dust attenuation estimate from \citet{2009ApJ...700..161S} is shown as a function of stellar mass for group galaxies at $0.8<z<1$ with secure redshifts. The {\it thin solid line} and {\it thin dotted line} represent our 50 per cent completeness limit due to the $M/L_r$ distribution at fixed sSFR, at $z=0.8$ and $z=1$, respectively. Galaxies with 24 micron MIPS detection within 3$''$ are outlined in purple. The star-forming sequence from \citet{2012ApJ...745..179W} {\it (think line)} at $z=0.9$ are included for comparison, along with the $4\times$ sSFR($z$) threshold {\it (think dashed line)} from \citet{2011ApJ...739L..40R}. Note that those points with upper limits are not detected in the FUV. {\it Right:} The same plot, but showing the results for the field sample.}
	\label{fig-sfrfuvir}
\end{figure*}
\begin{figure}
	\leavevmode
	\epsfysize=8.5cm
	\epsfbox{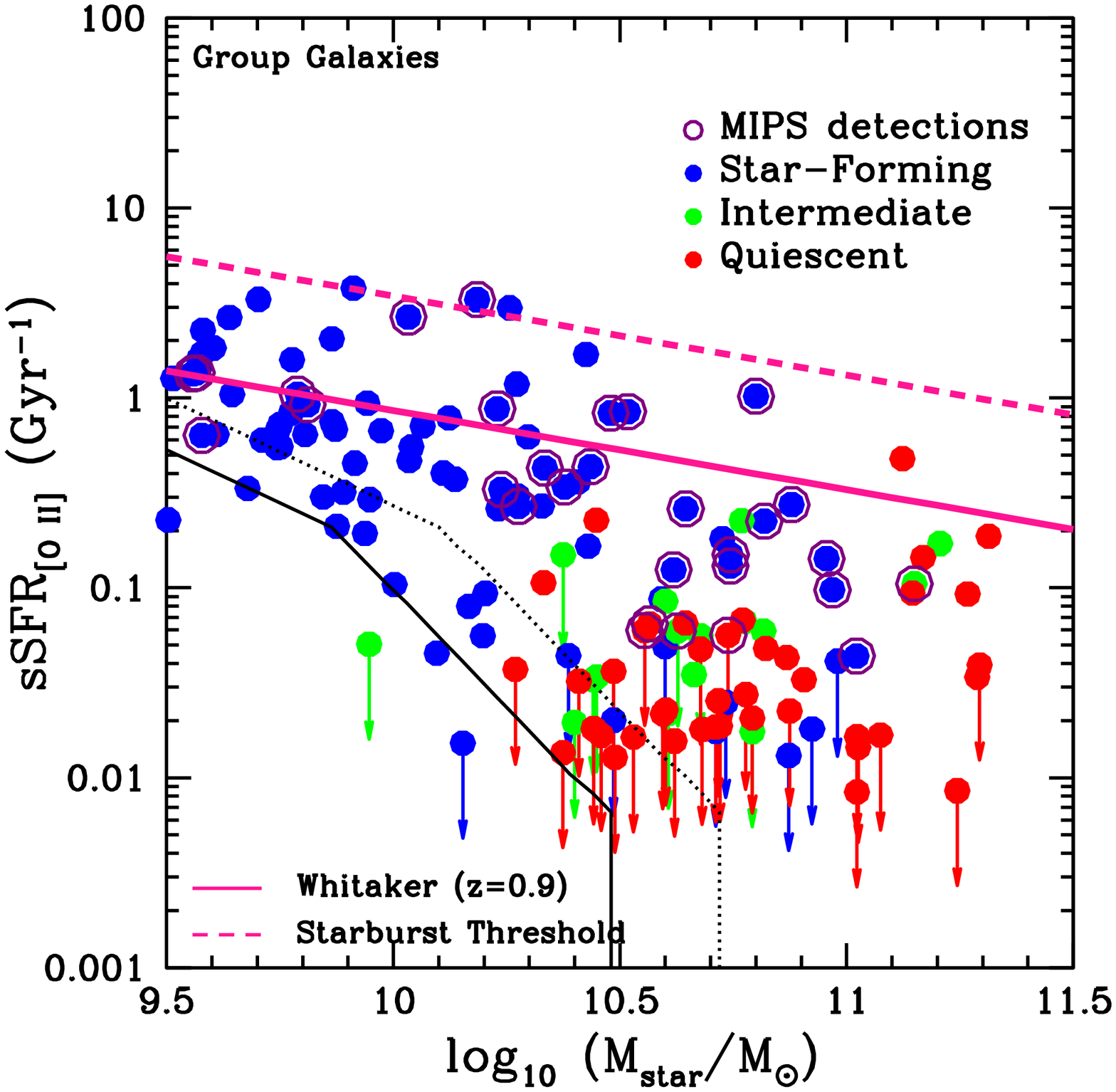}
	\caption{The specific star formation rate measured from the \oii feature is shown as a function of stellar mass for group galaxies at $0.8<z<1$ with secure redshifts. The {\it thin solid line} and {\it thin dotted line} represent our 50 per cent completeness limit due to the $M/L_r$ distribution at fixed sSFR, at $z=0.8$ and $z=1$, respectively. Galaxies with 24 micron MIPS detection within 3$''$ are outlined in purple. The star-forming sequence from \citet{2012ApJ...745..179W} {\it (think line)} at $z=0.9$ are included for comparison, along with the $4\times$ sSFR($z$) threshold {\it (think dashed line)} from \citet{2011ApJ...739L..40R}. Note that those galaxies with a negative \oii equivalent width or with \oii detections less than the measurement error are converted into upper limits.}
	\label{fig-sfroii}
\end{figure}
\par
The resulting plot of the specific star formation rate (sSFR) vs stellar mass for the FUV + IR and the FUV + dust attenuation estimates are presented in Figure~\ref{fig-sfrfuvir}. Those points undetected in the FUV are converted into upper limits with the survey detection limit of $25$ mag. We notice that most of the star-forming galaxies are detected in FUV, while many of the intermediate and most of the quiescent galaxies are not.
\par
The star-forming group galaxies agree with the star forming sequence for field galaxies at $z=0.9$ of \citet{2012ApJ...754L..29W}, as seen in Figure~\ref{fig-sfrfuvir}, especially where the GEEC2 sample is complete ($>10^{10.5} M_\odot$). From \citet{2013MNRAS.431.1090M}, we note that for intermediate galaxies, the average specific star formation rate inside groups is lower than in the field for our small sample.
\par
The star-forming group sample contain 90 galaxies in total. If we take the limit of $4\times$ sSFR($z$) of `normal' star-forming galaxies threshold from \citet{2011ApJ...739L..40R} to be the signature for galaxies displaying an excess in star formation, then only $4.4\substack{+3 \\ -1}$ per cent fit this criteria (4 out of 90). In the field, this value is $5.1\pm1$ per cent (44 out of 871). Although our sample is small, this result is in agreement with the findings of other groups which have not found an enhancement in star-formation in higher density regions \citep[e.g.][]{2013ApJ...770...62D, 2013arXiv1304.3335W}.
\par
Next, the results for the \oii - based star formation rate measurements are presented in Figure~\ref{fig-sfroii}. Overall, the properties of the star-forming galaxies remained roughly the same as from the previous method, but this method is able to measure lower star formation rates, as some of the objects are undetected in FUV but do have some \oii emission. If we again take the limit of $4\times$ sSFR($z$) as the threshold for an enhancement in star formation relative to the normal population, then only $6.7\substack{+4 \\ -2}$ per cent fit this criteria (6 out of 90). This matches the results from the FUV + IR measurements and shows that starbursts may not play a big role in environmental quenching. Since such enhancements do not appear to be the primary process in the transition between star-forming and quiescent states, this result motivates our use of a simple quenching model, where star formation is quenched after infalling into the group.
%
%
\section{Modelling the Intermediate Population}
\begin{figure}
	\leavevmode
	\epsfysize=8.5cm
	\epsfbox{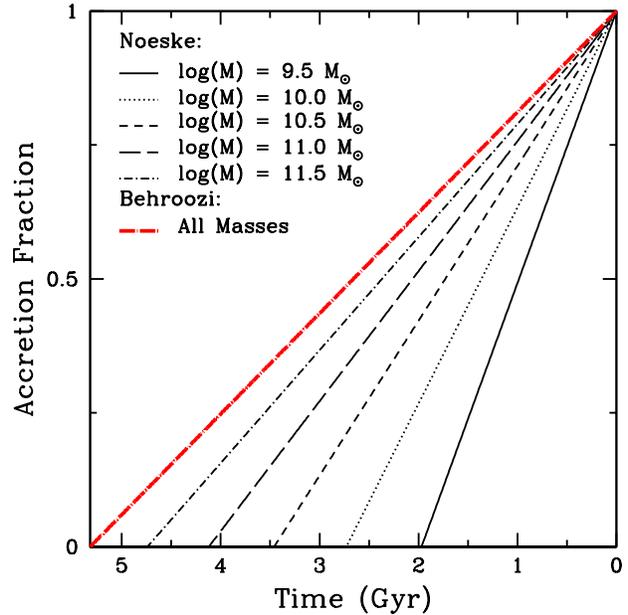}
	\caption{Sample accretion curves for different model galaxies with Noeske's star formation histories. Galaxies with stellar masses of $10^{9.5}M_\odot$, $10^{10.0}M_\odot$, $10^{10.5}M_\odot$, $10^{11.0}M_\odot$, and $10^{11.5}M_\odot$ are presented. Also included is the mass-independent accretion curve used for the galaxies with Behroozi's star-formation histories.}
	\label{fig-infall}
\end{figure}%
\subsection{Stellar Synthesis Models}
\par
The stellar synthesis code in this paper uses the \citet{CB07} models (also known as CB07), an update to the GALAXEV BC03 code \citep{2003MNRAS.344.1000B}. This program takes in a user-defined star formation history and outputs the model galaxy's physical characteristics throughout its lifetime, such as its flux through different filters, its resulting spectra, as well as other important data. The \hdelta strength can also be measured in the model synthesis code, since it uses the bandpass method in an similar wavelength range to our analysis in \citet{2013MNRAS.431.1090M}.
\par
The observations can be compared to the model results and used to constrain the possible star formation histories of the galaxies in the GEEC2 survey. Given the low signal-to-noise of spectra at $z\sim1$, this method is best used on bins of galaxies with similar stellar mass, colour and environment. For this paper, we will be looking at the collective properties of six main types of galaxies: quiescent group, quiescent field, intermediate group, intermediate field, star-forming group, and star-forming field.
\par
Finally, the classification system used in the GEEC2 survey can be reproduced for these model galaxies using the CB07 code. The $(V-z)^{0.9}$ and $(J-[3.6])^{0.9})$ colours, used in the GEEC2 survey, can be recreated. This process is done by red-shifting the $V$, $z$, $J$, and $[3.6]$ filter response functions to $z=0.9$. The two colours can then be calculated and assigned to these model galaxies. Our choice of the colour cuts to separate model galaxies into quiescent, intermediate, and star-forming will be discussed in \S~\ref{sec-colourcuts}.
\subsection{Satellite Quenching Model}
\par
Once the framework is in place, we need to create standard star-formation histories for these quenched satellites. If we assume that these objects were `blue' before entering the group environment, then we need to create sample star-formation histories for actively star forming objects. This assumption may be relatively safe, given the average infall times and the fraction of star-forming galaxies at these high redshifts \citep{2013arXiv1303.4409M}.
\par
In this paper, two different models are used. First, the star formation model from \citet{2012arXiv1207.6105B} is assumed. They used cosmological models and fit models for the stellar mass-halo mass relation to determine the most likely star formation histories for different mass haloes. Using the data set found at \url{http://www.peterbehroozi.com/data.html}, the closest star formation history for haloes that produce central galaxies with stellar masses of $10^{9.5}$, $10^{10.0}$, $10^{10.5}$, $10^{11.0}$ and $10^{11.5}$ $M_\odot$ were chosen.
\par
For comparison, the staged star formation model of \citet{2007ApJ...660L..47N} is also used, where they have fitted to the observed distribution of specific star formation rates (sSFR) at $z=0$ with an empirical function. In the model, the star formation history of a galaxy is an exponential function, with parameters (such as the formation times and exponential timescale $\tau$) that will be dependent on the galaxy's mass. Except for the observation time, which is set to $z=0.9$, all other parameters are kept as in the original paper. Note that in general for Noeske's model, lower mass objects have their major period of star formation start later, possess a longer $\tau$, and have a higher specific star formation rates at $z=0.9$. Conversely, higher mass objects have their major period of star formation start earlier, possess a shorter $\tau$, and have a lower specific star formation rates at $z=0.9$.
\begin{figure*}
	\leavevmode
	\epsfysize=8.5cm
	\epsfbox{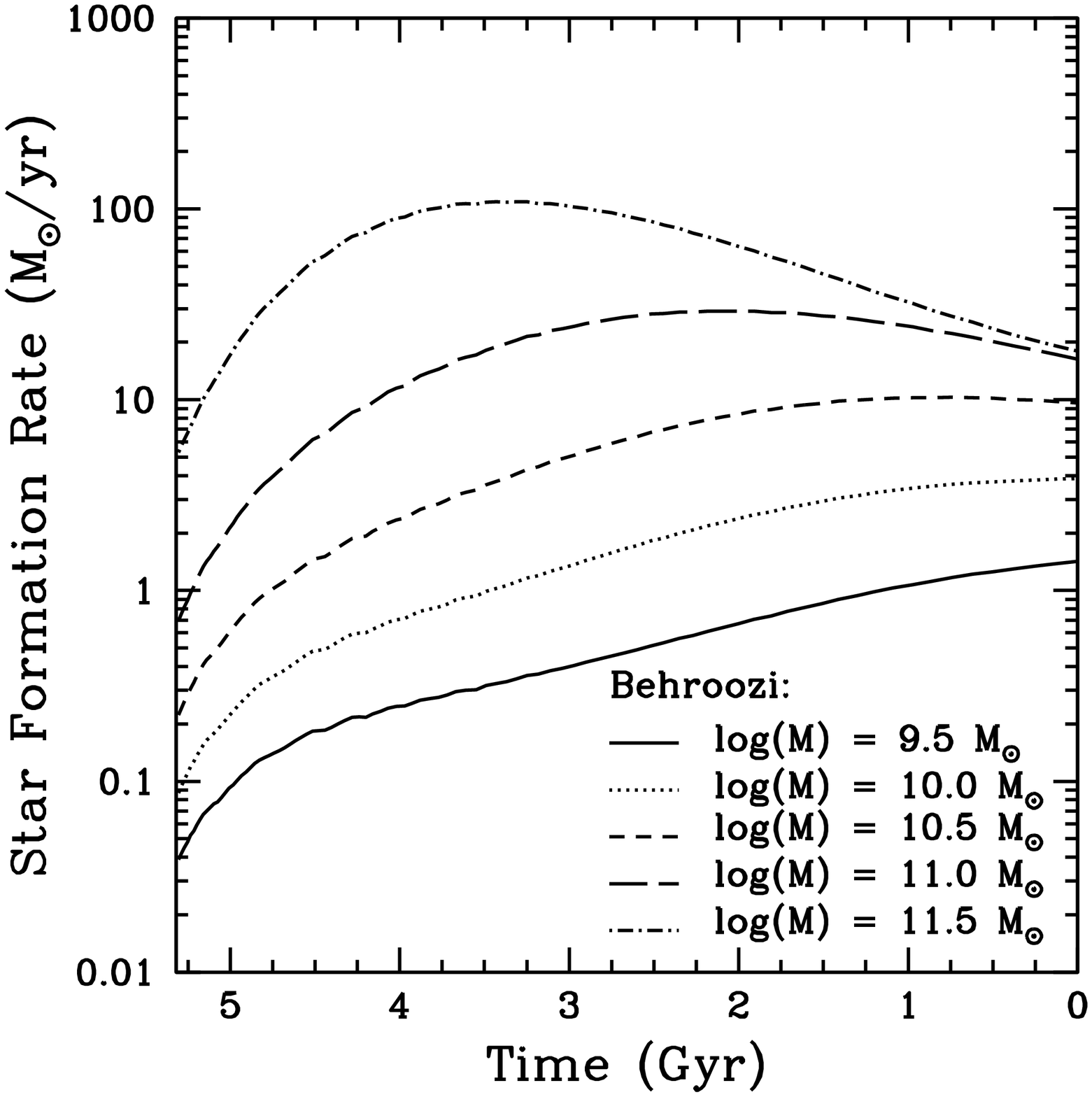}
	\epsfysize=8.5cm
	\epsfbox{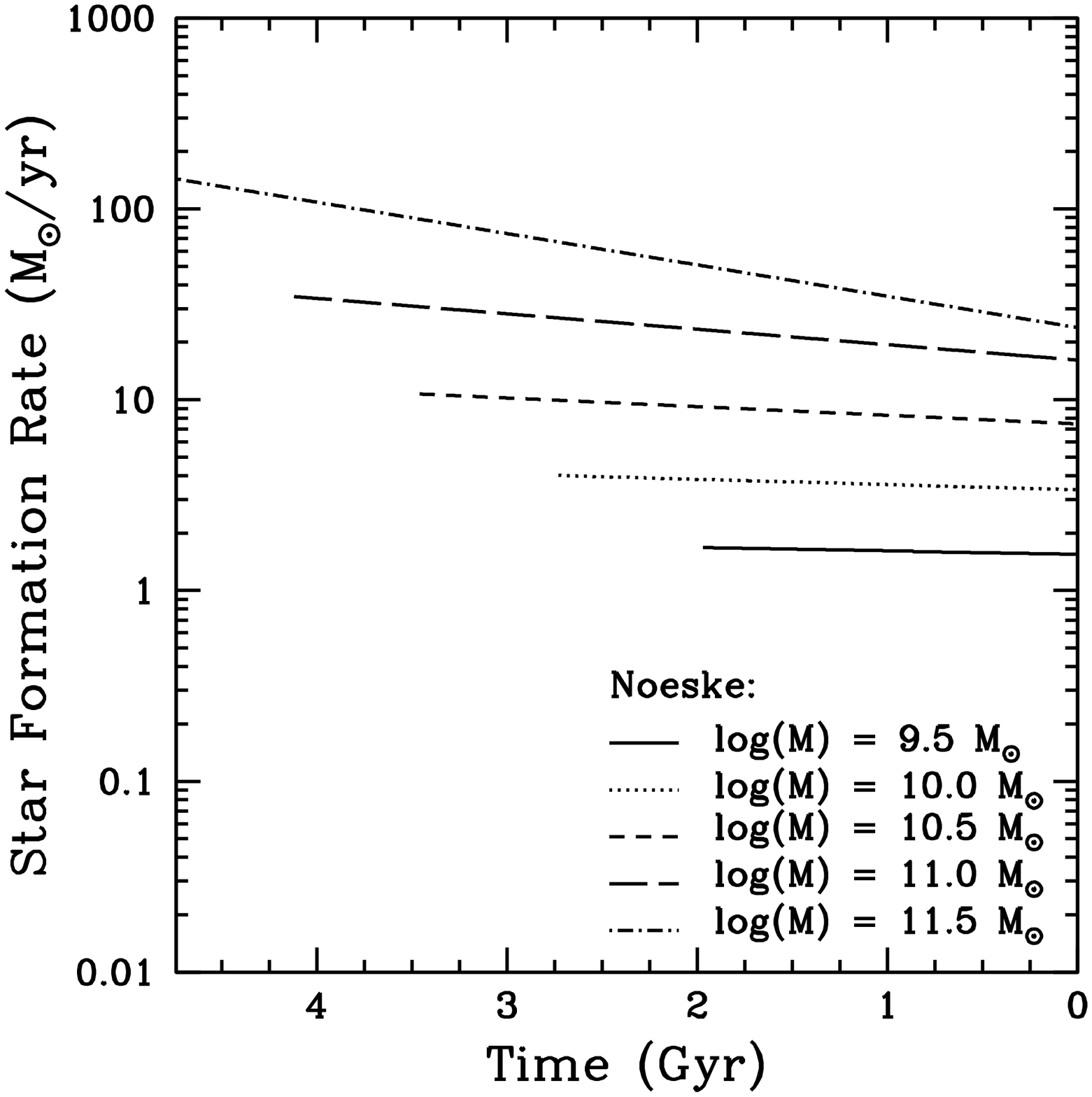}
	\caption{{\it Left:} The star formation rate histories for galaxies with stellar masses of $10^{9.5}M_\odot$, $10^{10.0}M_\odot$, $10^{10.5}M_\odot$, $10^{11.0}M_\odot$, and $10^{11.5}M_\odot$, taken from the \citet{2012arXiv1207.6105B} models. Note that the lower mass objects have increasing star formation rates up to the time of observation, while the higher mass models have already peaked. {\it Right:} The star formation rate histories for the galaxies with stellar masses of $10^{9.5}M_\odot$, $10^{10.0}M_\odot$, $10^{10.5}M_\odot$, $10^{11.0}M_\odot$, and $10^{11.5}M_\odot$, taken from the \citet{2007ApJ...660L..47N} models. Note that galaxies of different stellar masses have different formation times and exponential timescales ($\tau$).}
	\label{fig-sfrh}
\end{figure*}
\subsubsection{Satellite Infall Rates}
\par
The semi-analytic galaxy formation models of \citet{2006MNRAS.370..645B} for groups at this redshift ($z\sim1$) indicate that the infall rate should roughly be constant for $\sim6$ Gyr before $z\sim1$. We will therefore assume that there is a uniform distribution of infall times between the formation time of the satellite galaxy and the time of observation at $z=0.9$. This amount of time varies for Noeske's model, which have different formation times for different mass galaxies. However, all of Behroozi's models start at $z=7.95$ and galaxies of different masses would have the same formation time.
\par
The satellite accretion histories are normalized such that all group galaxies are assumed to have accreted by the time of observation. In all cases, the accreted fraction rises from 0 at the time of formation to 1 at $z=0.9$. In Noeske's model, a lower mass object is `formed' later and thus would have a steeper accretion curve, in order to have been observed in the group environment at this redshift. The dependence of the slope of the accretion curve and the mass of the satellite galaxy can be seen in Figure~\ref{fig-infall}.
\subsubsection{Procedure}
\par
Using the \citet{2012arXiv1207.6105B} star formation histories and the \citet{2007ApJ...660L..47N} staged galaxy formation model, we produce star formation histories for galaxies with stellar masses of $10^{9.5}$, $10^{10.0}$, $10^{10.5}$,  $10^{11.0}$, and $10^{11.5}$ $M_\odot$ at the time of observation. This is shown in the left and right plot in Figure~\ref{fig-sfrh} for the Behroozi and Noeske star formation histories respectively. Note that for their presentation in this plot, the star-formation histories have been normalized such that the integral of their star formation rate from their formation to the time of observation is equal to their stated stellar mass.
\par
Next, the `natural' rate of star formation is modified once a galaxy has entered the group. We assume that the satellite galaxy undergoes a `quenching' process, with an exponential timescale ($\tau_Q$), where an immediate cut-off in star formation can be modelled with a $\tau_Q=0$ Gyr. In addition, this process can be modified by the addition of a delay before quenching begins, as advocated by \citet{2012MNRAS.424..232W}. These changes can have a large effect on the fractions of quiescent, intermediate, and star-forming galaxies, as well as the strengths of their \hdelta feature. As a result, the different choices of quenching parameters can be constrained by our observations.
\par
A set of models is created with a range of quenching start times, spaced out at time intervals of 0.02 Gyr from the galaxy's formation time to the time of observation. Each individual model can be identified with a specific point on the accretion curve, simulated independently, and combined for our final results.
\par
Note that the results from the stellar synthesis code are indifferent to the absolute values from the star-formation histories, only the overall shape. Therefore, for these individual star formation histories, we use the unquenched models for each stellar mass satellite galaxy as the benchmark. The stellar masses at the end of the models (excluding stellar remnants), are then normalized with the same factor, such that the stellar mass for the unquenched model would be at $10^{9.5}$, $10^{10.0}$, $10^{10.5}$, $10^{11.0}$, and $10^{11.5}$ $M_\odot$ respectively.
\par
By keeping track of the resulting stellar mass of these models at the time of observation, these objects would then be matched to the correct mass bin. We choose bins of equal sizes in log space, for example, $10^{9.5}$ $M_\odot$ would correspond to galaxies with a stellar mass of $10^{9.25}$ $M_\odot$ to $10^{9.75}$ $M_\odot$. This is necessary because for some extreme cases, the `earliest' quenched objects may not have enough stellar mass to remain in their original mass bins. We use the stellar mass functions for star-forming group galaxies from \citet{2012A&A...538A.104G} to determine the relative proportion of initial objects at each mass bin and weight accordingly. 
\par
Then, we run the BC07 models and calculate the weighted average of the resulting colour and spectral features at the time of observation, in order to determine the general properties of these populations. Note that this procedure assumes that all the group galaxies would be star-forming at the time of accretion, as we use star formation histories for actively star forming galaxies as the initial inputs. This may not be inappropriate assumption, given the infall times required to be observed as a group member at $z\sim1$ and the previously observed trends in the stellar mass function of star-forming and quiescent galaxies with redshift \citep{2013arXiv1303.4409M}.
\par
In addition, if these group galaxies transition to a passive state through a process not related to environment (such as through Peng's mass-quenching), then this model may over-predict the number of star-forming (`blue') galaxies. Finally, we have not modelled galaxies that have temporarily shifted from the star-forming to the intermediate stage due to changes in the gas accretion rate. As a result, the constraints on $\tau_Q$ from the intermediate fraction may be even stronger.
\subsection{Choice of Model Colour Cuts}\label{sec-colourcuts}
\begin{figure}
	\leavevmode
	\epsfysize=8.5cm
	\epsfbox{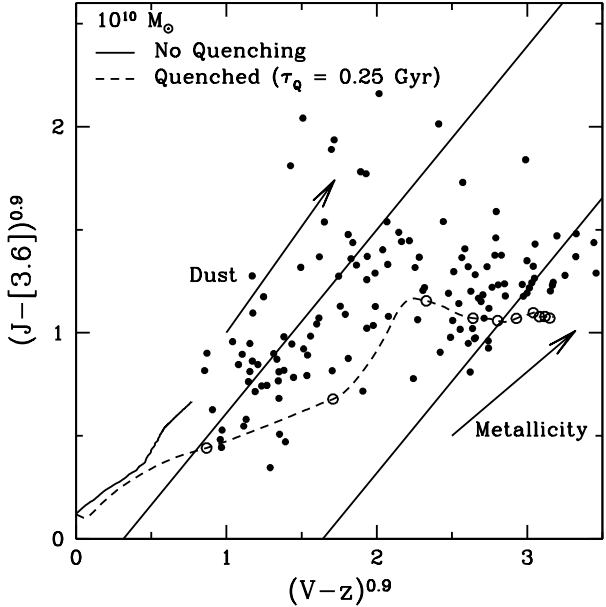}
	\caption{Model tracks for a star-forming galaxy in the Behroozi model (with a stellar mass of $10^{10} M_\odot$, no dust extinction, and a solar metallicity $Z=0.02$) vs a sample quenched model with $\tau_Q=0.25$ Gyr in the $(J-[3.6])^{0.9})$ vs $(V-z)^{0.9}$ plane. The group galaxies from the GEEC2 survey are presented as {\it black filled circles}. The {\it solid lines} show the GEEC2 fit to the star-forming and quiescent galaxy distribution. {\it Black circles} show time intervals of 0.50 Gyr for the quenched model. Also plotted are the dust (from $\tau_v = 0$ to $\tau_v = 5.0$) and metallicity arrows (from $Z=0.004$ to $Z=0.08$). We see that the data points can be spanned with a combination of these simple models and the contribution from dust and metallicity.}
	\label{fig-dustmetal}
\end{figure}
\begin{figure*}
	\leavevmode
	\epsfysize=8.5cm
	\epsfbox{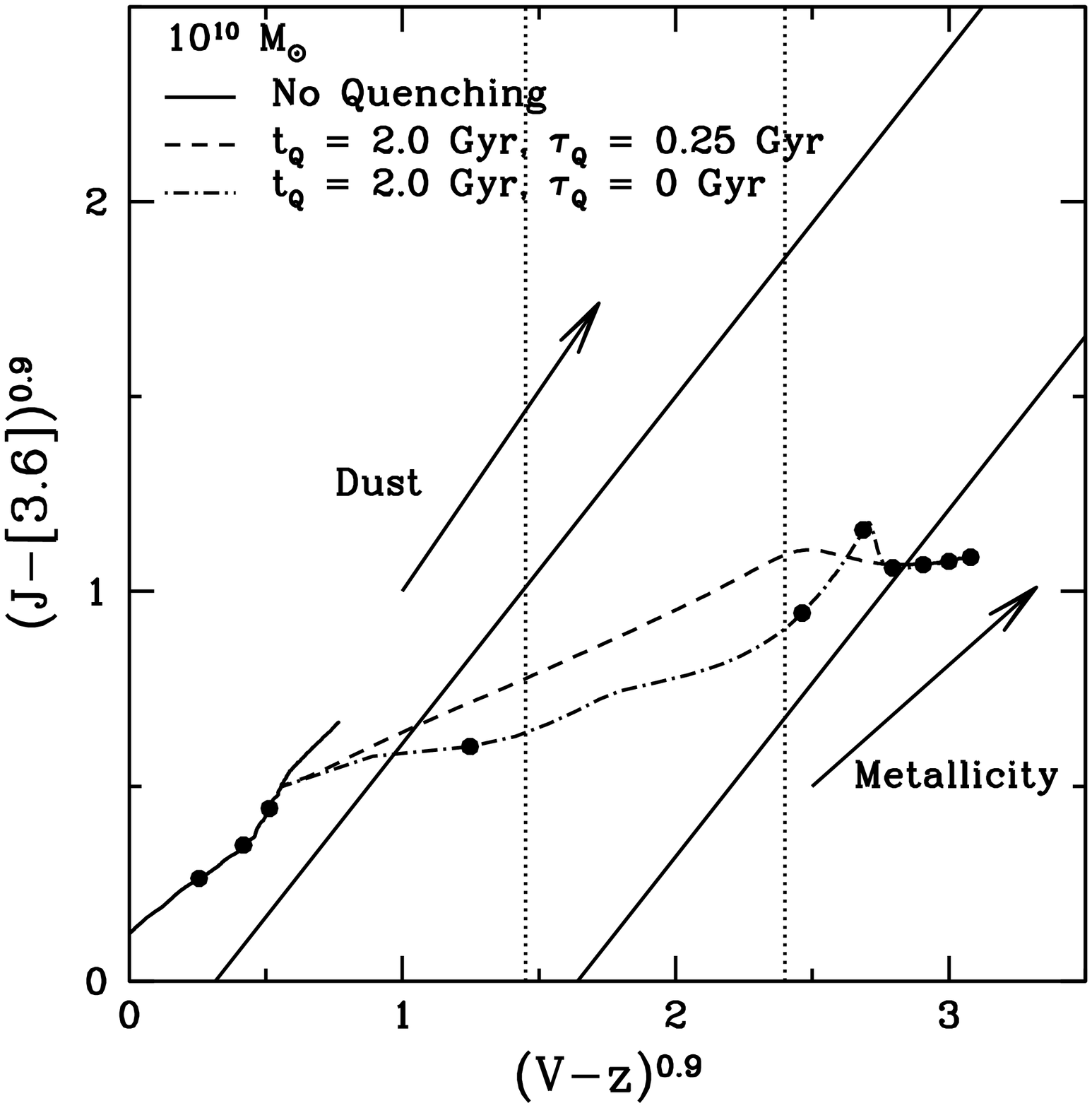}
	\epsfysize=8.5cm
	\epsfbox{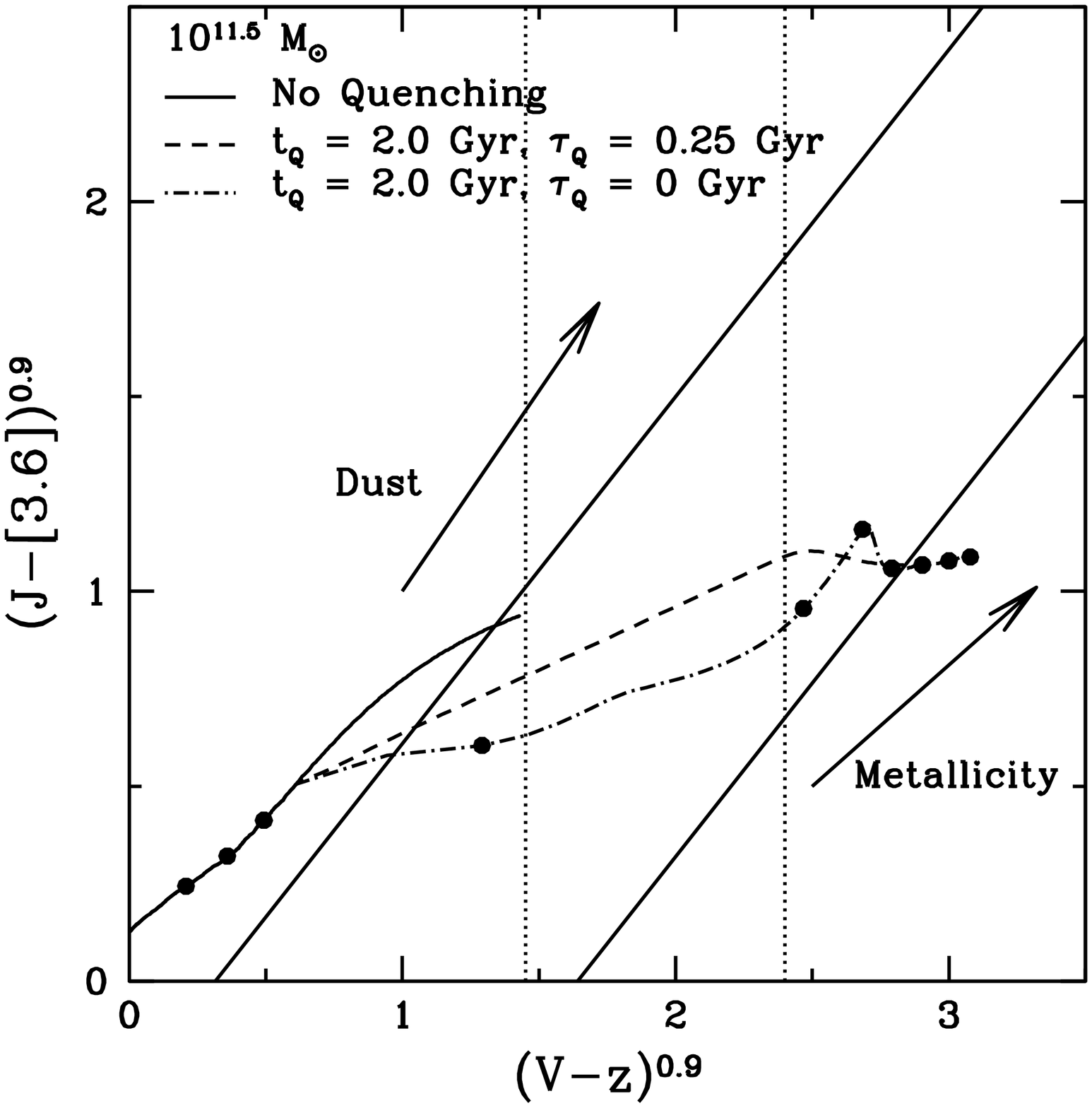}
	\caption{{\it Left:} The effects of the quenching process for a sample galaxy with the Behroozi models, a stellar mass of $10^{10.0}M_\odot$, and the starting time for the quenching process (t$_Q$) at 2.0 Gyr after formation. The model track is shown in the $(J-[3.6])^{0.9})$ vs $(V-z)^{0.9}$ plane, with no dust and solar metallicity. The quenching timescale ($\tau_Q$) is set at 0 Gyr (immediate cutoff) and 0.25 Gyr. Also plotted are the average dust (from $\tau_v = 0$ to $\tau_v = 5.0$) and metallicity arrows (from $Z=0.004$ to $Z=0.08$). {\it Black dots} show time intervals of 0.50 Gyr for the immediate cut-off model. The {\it solid lines} show the GEEC2 fit to the star-forming and quiescent galaxy distribution. {\it Right:} The effects of the same quenching process on a sample galaxy, but with a stellar mass of $10^{11.5}M_\odot$.}
	\label{fig-bound}
\end{figure*}
\par
To directly compare between the model galaxies and the populations of the GEEC2 survey, we need to determine the best way of separating the quiescent, intermediate, and star-forming populations. While we can retain the definitions from the GEEC2 survey, there is a more efficient method in terms of the number of models required to be run. This new method can reduce the large number of parameters to be explored, such as quenching parameters, stellar mass, dust, and metallicity.
\par
We first test the effects of dust and metallicity, which were the primary motivations behind the classification of \citet{2013MNRAS.431.1090M}. We follow the track of an instantaneous burst model through the $(J-[3.6])^{0.9}$ vs $(V-z)^{0.9}$ plane, while varying the model parameters for the dust and metallicity. This is shown as the dust and metallicity arrows in Figure~\ref{fig-dustmetal}. The stellar synthesis models have dust and metallicities vectors that are roughly parallel to the star-forming sequence used to separate our galaxies in the GEEC2 sample. 
\par
As we have anticipated, our original procedure for the separation into quiescent, intermediate, and star-forming galaxies has largely incorporated the effect of dust and metallicity. The star-forming sequence and the `red' sequence can be reproduced by our star-forming model and the quenched model, as shown overlaid on the group galaxies in Figure~\ref{fig-dustmetal}. The intermediate galaxies are found in between the two main distributions. Note that some of the star-forming galaxies may require an extinction of up to $\tau_v\sim4-5$, which also roughly corresponds to the region where many MIPS detected galaxies are found and where such high extinctions have been observed.
\par
Since the effect of dust and metallicity would only move the galaxies diagonally along the two sequences, we can collapse the colour classification onto the x-axis, $(V-z)^{0.9}$. This can be accomplished by projecting all galaxies along the diagonal lines corresponding to the possible action of dust and/or metallicity. Therefore, we can create one set of model galaxies with no dust and solar metallicity content while still maintaining a good correspondence to our previous colour classifications.
\par
To find out the required classifications along the $(V-z)^{0.9}$ axis, we first plot the model tracks for different stellar mass galaxies, with the two extreme cases of $10^{10.0}$ and $10^{11.5}$ $M_\odot$. This is shown in Figure~\ref{fig-bound}.
\par
For the Behroozi models, the highest mass star-forming models have already peaked in star formation by $z\sim1$. These galaxies would have evolved more than for the lower mass case, which have increasing star formation rates at $z\sim1$, as seen in Figure~\ref{fig-sfrh}. The boundary for the Behroozi case is therefore set at $(V-z)^{0.9}=1.45$ and any galaxy with a redder colour would be considered to be intermediate (`green') or quiescent objects. This boundary also works well for the Noeske models, as this is also close to the limit for the highest mass and longest evolving Noeske star-forming models.
\par
Next, we want to set the approximate boundary between the intermediate (and potentially transitioning) population and the quiescent population. We can follow the track of sample model galaxies after they have been environmentally quenched, as shown in Figure~\ref{fig-bound} for a gentler truncation ($\tau_Q=0.25$ Gyr) and an immediate cut-off model ($\tau_Q=0$ Gyr).
\par
The result shows that the quenching process takes the galaxies quickly from the star-forming (`blue') to the quiescent (`red') sequence, from the dots indicating the time intervals of 0.50 Gyr. We have set the boundary between the quiescent and intermediate population at $(V-z)^{0.9}=2.4$, the point at which the evolution of these quenched model noticeably slows down, as shown in Figure~\ref{fig-bound}. Note that the exact location of these two boundaries does not significantly affect our results.
%
%
\section{Results}
\subsection{Satellite Quenching Models with No Delay}
\par
\begin{figure*}
	\leavevmode
	\rotatebox{90}{\parbox[c][1cm]{5.5cm}{\centering \Large No Delay (t$_Q=0$)}}
	\epsfysize=5.5cm
	\epsfbox{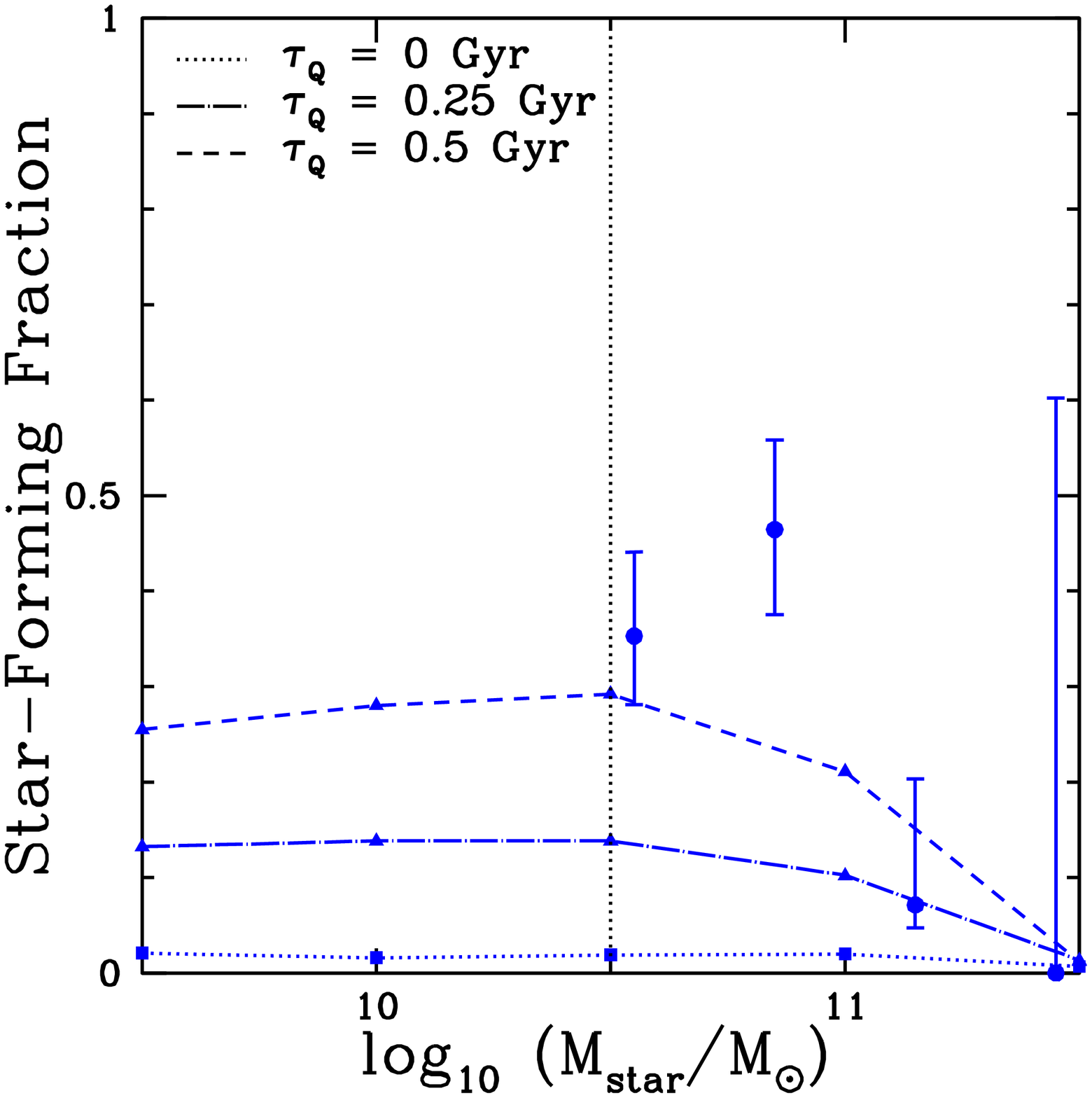}
	\epsfysize=5.5cm
	\epsfbox{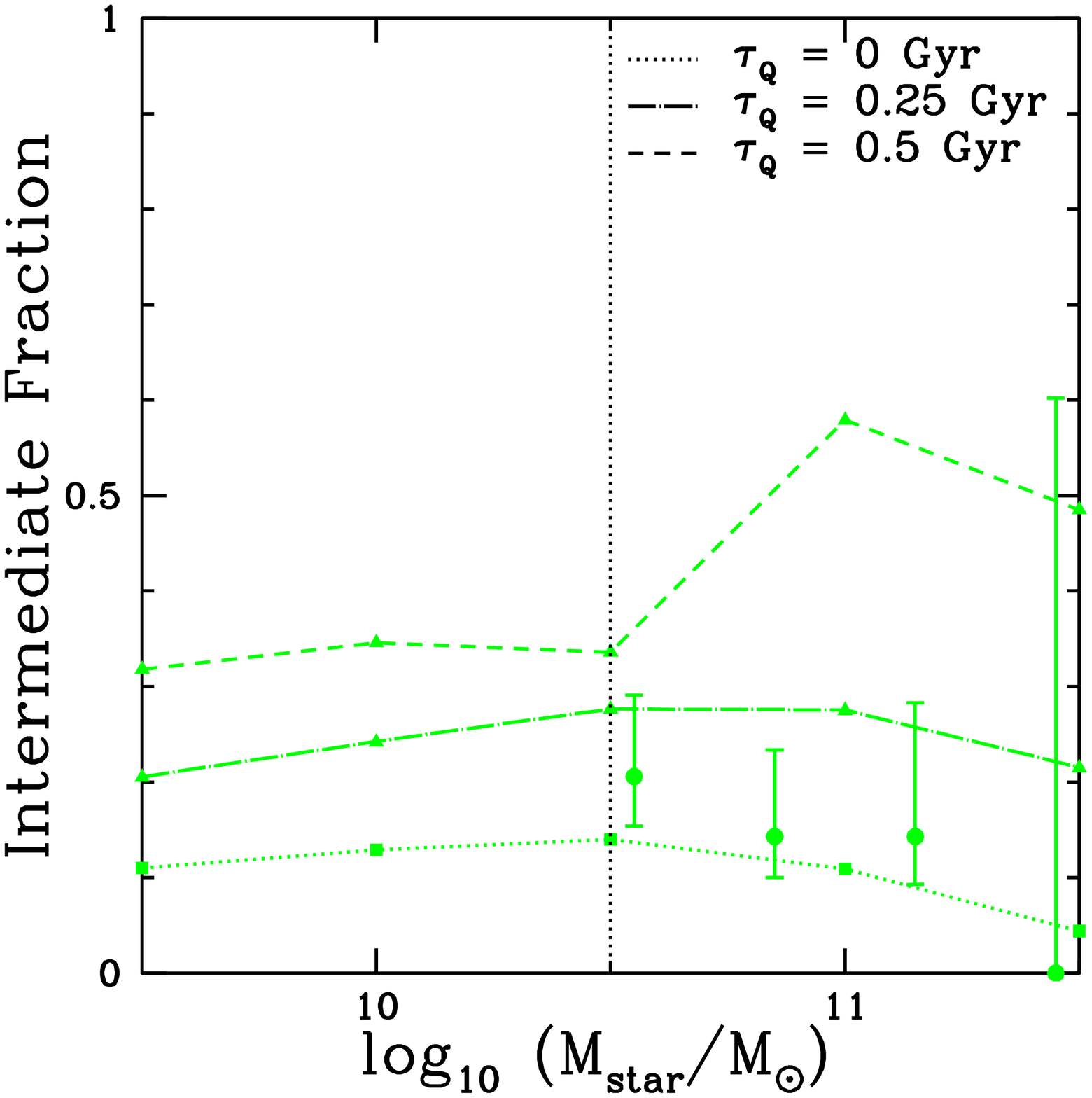}
	\epsfysize=5.5cm
	\epsfbox{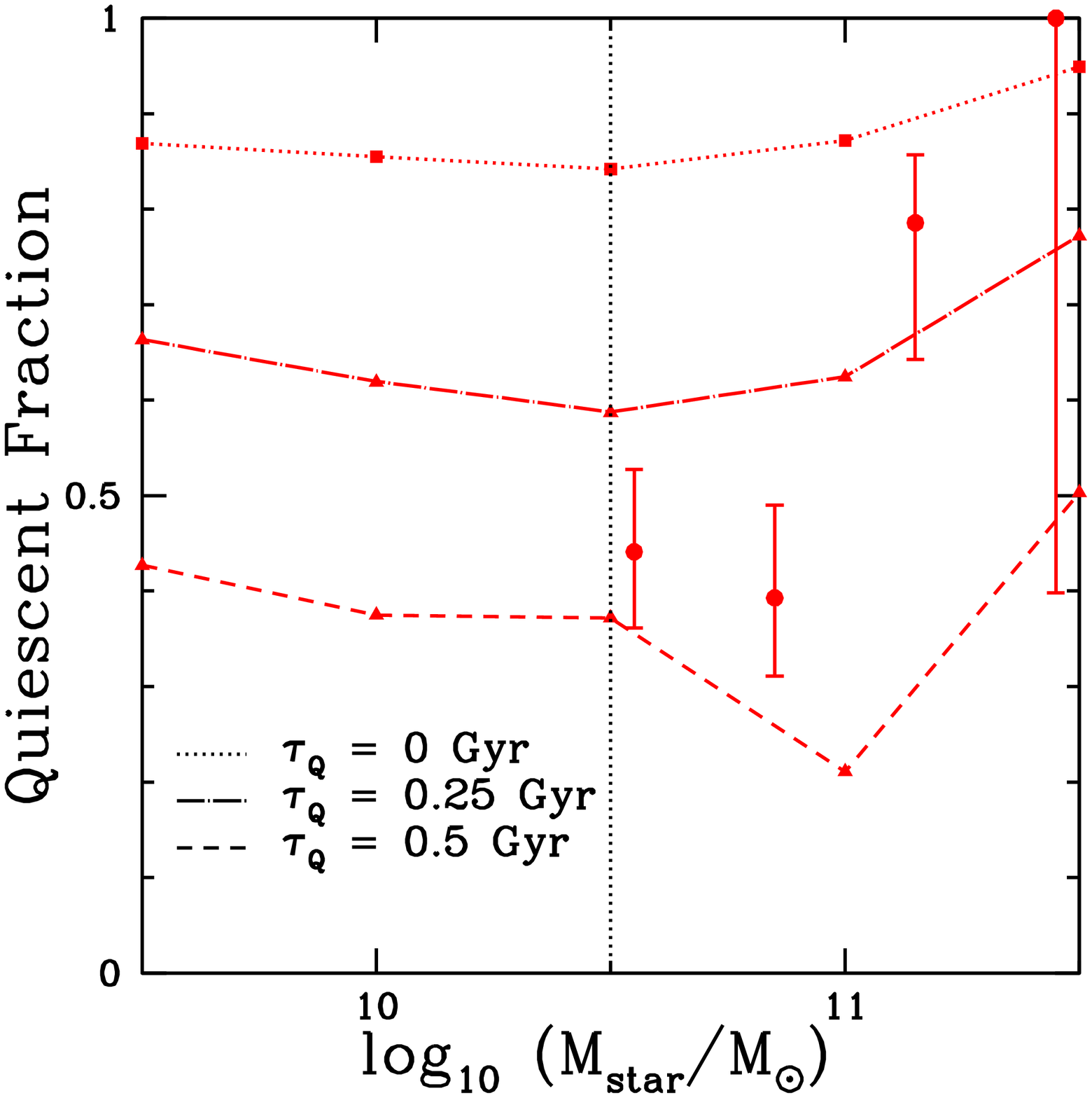}
	\caption{Satellite quenching model with varying $\tau_Q$ + {\bf no delay} and its impact on the star-forming {\it (left)}, intermediate {\it (centre)}, and quiescent {\it (right)} galaxy fractions for the Behroozi models. The {\it filled triangles} indicate different mass bins, with {\it dotted line} connecting models with $\tau_Q=0$ Gyr (immediate cut-off), {\it thick dashed line} connecting models with $\tau_Q=0.25$ Gyr, and {\it thin dashed lines} connecting models with $\tau_Q=0.5$ Gyr. The observed group fractions from GEEC2 are presented as points with error bars. The black dotted line at $10^{10.5} M_\odot$ shows where the sample is reasonably complete, even for passive galaxies.}
	\label{fig-Model_NoDelay_Frac}
\end{figure*}

To begin with, we assume a simple satellite quenching model for the Behroozi models, where the galaxy switches over to a quenched function with an exponential timescale ($\tau_Q$) immediately after accretion. In this case, three values for the quenched exponential timescale are chosen, $\tau_Q=0$ Gyr (immediate cut-off), $\tau_Q=0.25$ Gyr, and $\tau_Q=0.5$ Gyr. The resulting fractions of quiescent, intermediate and star-forming galaxies of this model are presented in Figure~\ref{fig-Model_NoDelay_Frac}.
\par
These models tend to under-produce star-forming galaxies, even for the longest $\tau_Q=0.5$ Gyr. The quenching process acts very quickly, as compared to the lifetime of the satellite galaxies. If there is no delay, then all galaxies accreted would quickly enter into the `green' phase and then become quiescent.
\par
The proportion of intermediate galaxies depend most strongly on the value of $\tau_Q$, as we have seen from plots of the model tracks in Figure~\ref{fig-bound}. An immediate cut-off model moves through the `green' region in a rapid fashion, compared to a less rapid truncation, and we would therefore expect to observe a small number of these intermediate (`green') galaxies. It is not possible to simultaneously match the proportion of star-forming/quiescent galaxies and the immediate-colour (`green') fraction, as the $\tau_Q=0.5$ Gyr model already produces an excess of `green' galaxies and still not enough star-forming galaxies.
\subsection{Satellite Quenching Model with Quenching Delay}
\begin{figure*}
	\leavevmode
	\rotatebox{90}{\parbox[c][1cm]{5.5cm}{\centering \Large t$_Q=3$ Gyr}}
	\epsfysize=5.5cm
	\epsfbox{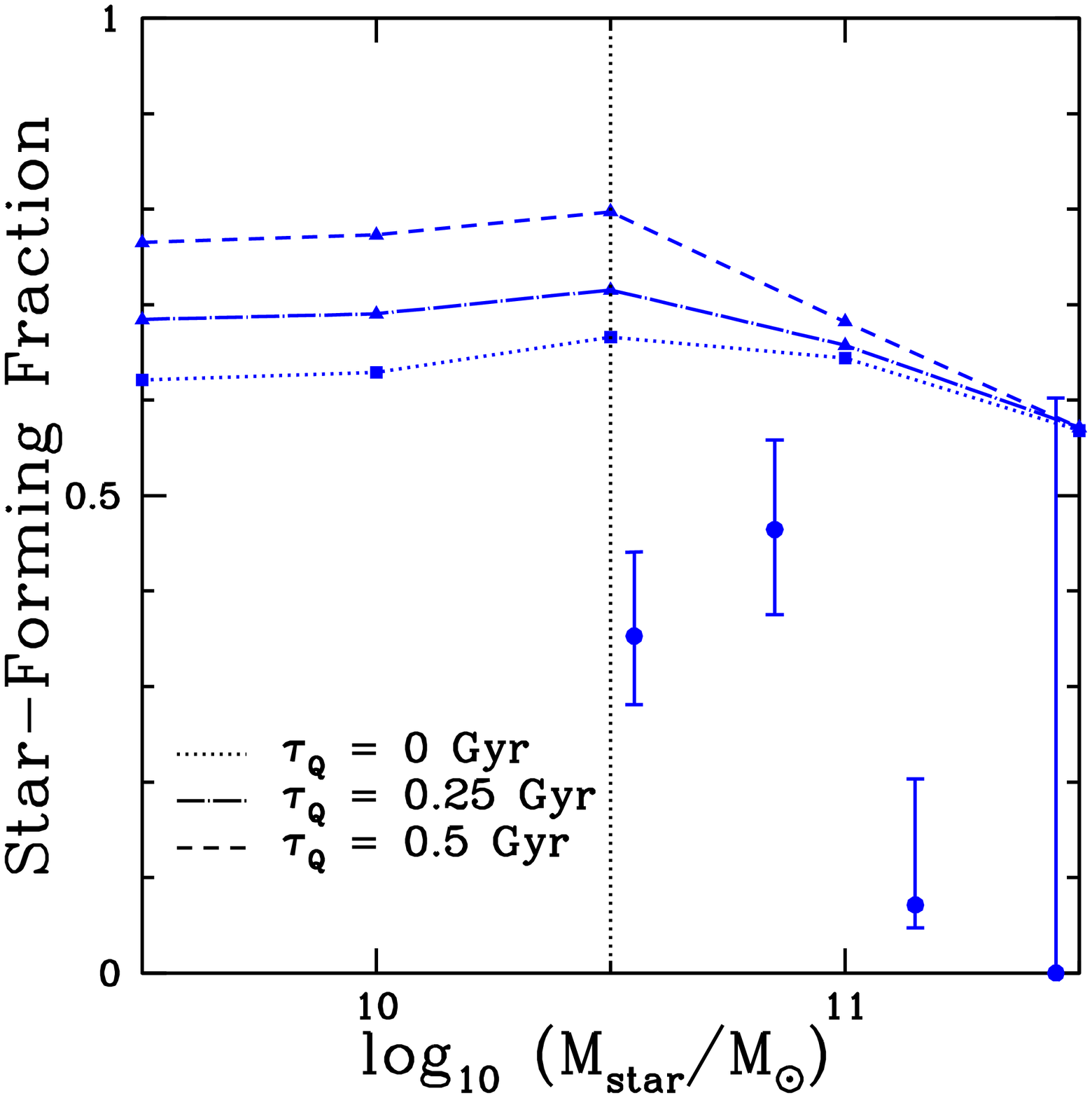}
	\epsfysize=5.5cm
	\epsfbox{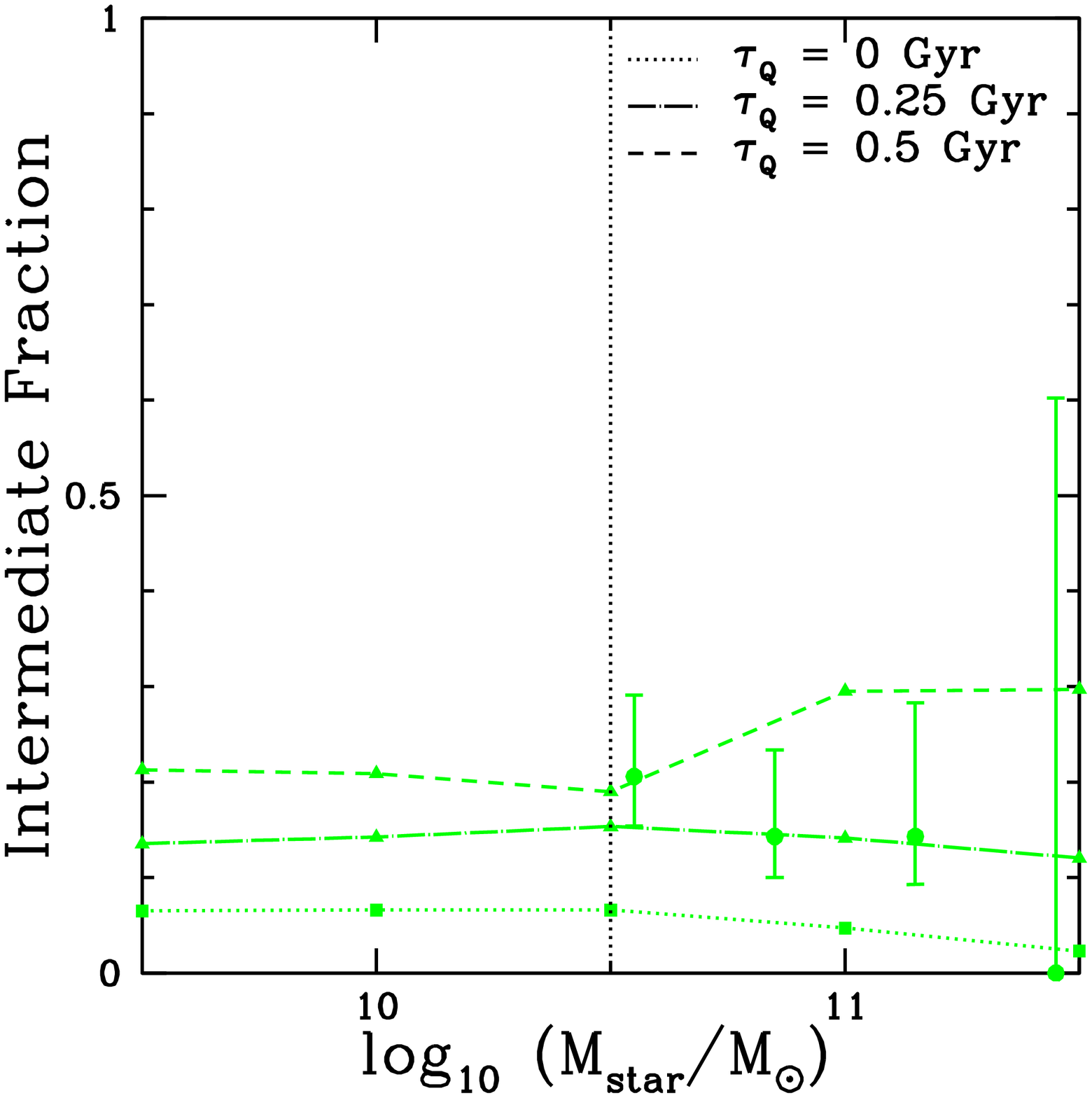}
	\epsfysize=5.5cm
	\epsfbox{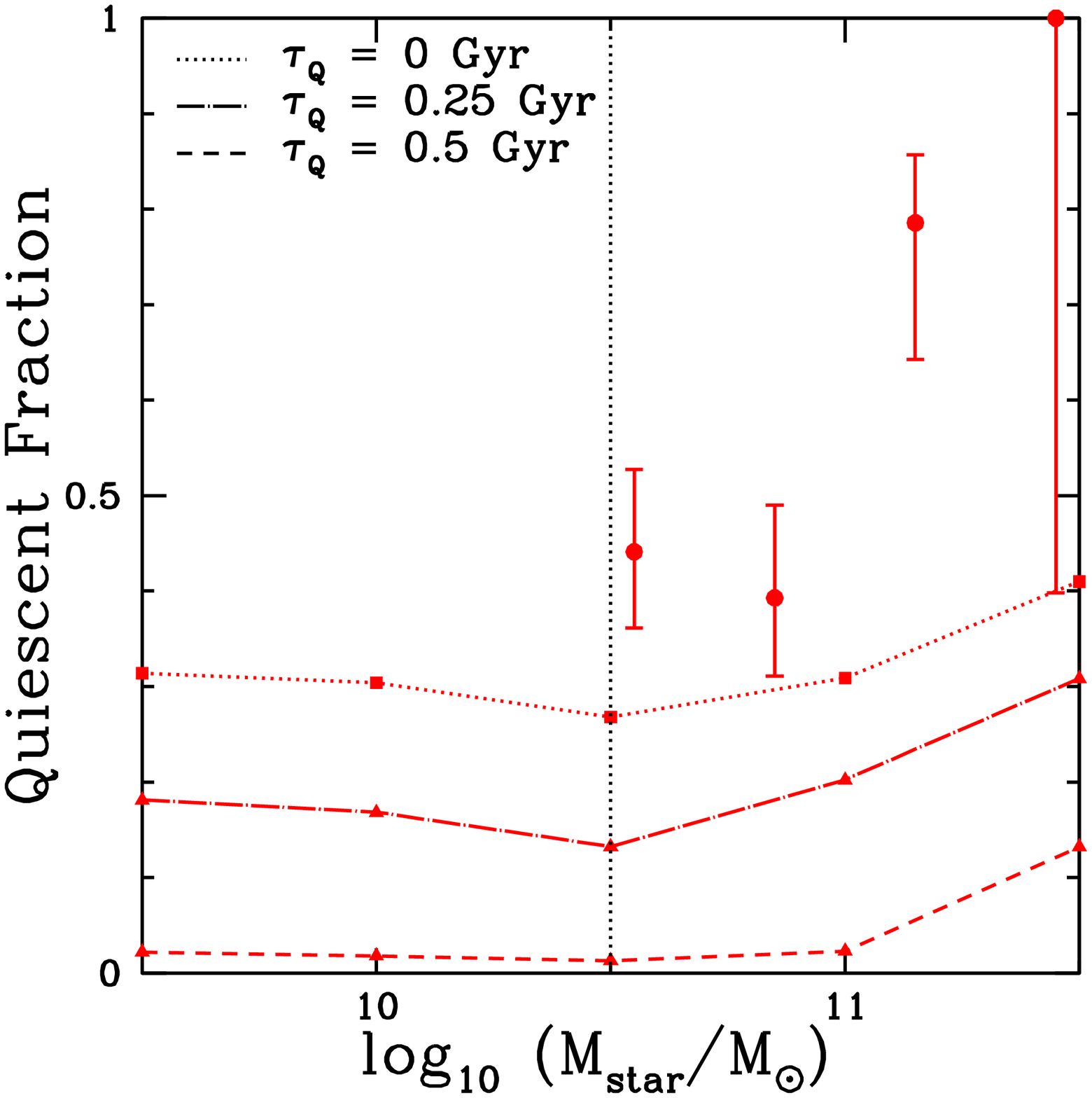}
	\caption{Satellite quenching model with varying $\tau_Q$ + {\bf a 3 Gyr delay} and its impact on the star-forming {\it (left)}, intermediate {\it (centre)}, and quiescent {\it (right)} galaxy fractions for the Behroozi models. The {\it filled triangles} indicate different mass bins, with {\it dotted line} connecting models with $\tau_Q=0$ Gyr, {\it thick dashed line} connecting models with $\tau_Q=0.25$ Gyr, and {\it thin dashed lines} connecting models with $\tau_Q=0.5$ Gyr. The observed group fractions from GEEC2 are presented as points with error bars. The black dotted line at $10^{10.5} M_\odot$ shows where the sample is reasonably complete, even for passive galaxies.}
	\label{fig-Model_3Delay_Frac}
\end{figure*}
\begin{figure*}
	\leavevmode
	\rotatebox{90}{\parbox[c][1cm]{5.5cm}{\centering \Large t$_Q=3(1+z_{\rm infall})^{-\frac{3}{2}}$}}
	\epsfysize=5.5cm
	\epsfbox{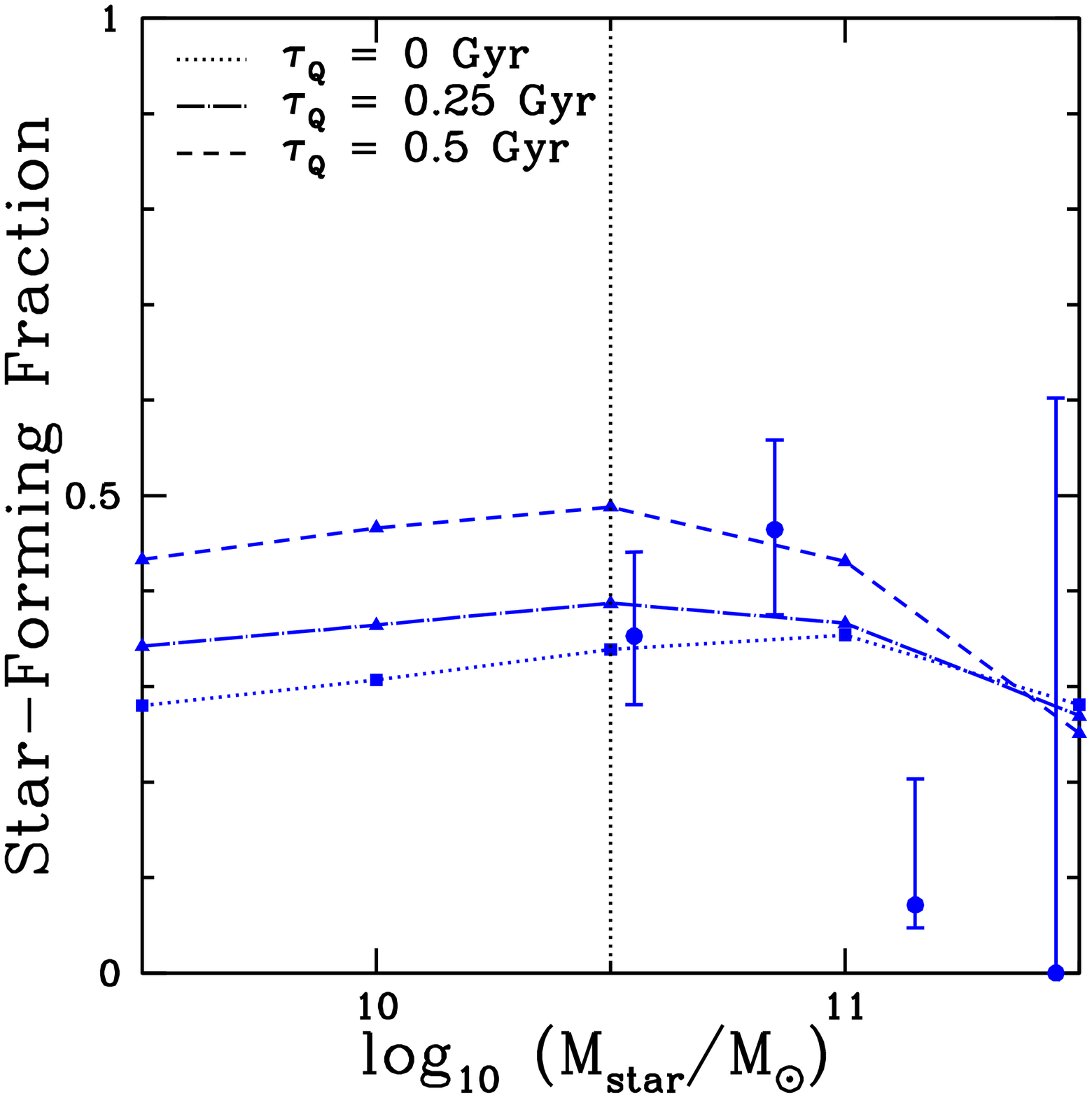}
	\epsfysize=5.5cm
	\epsfbox{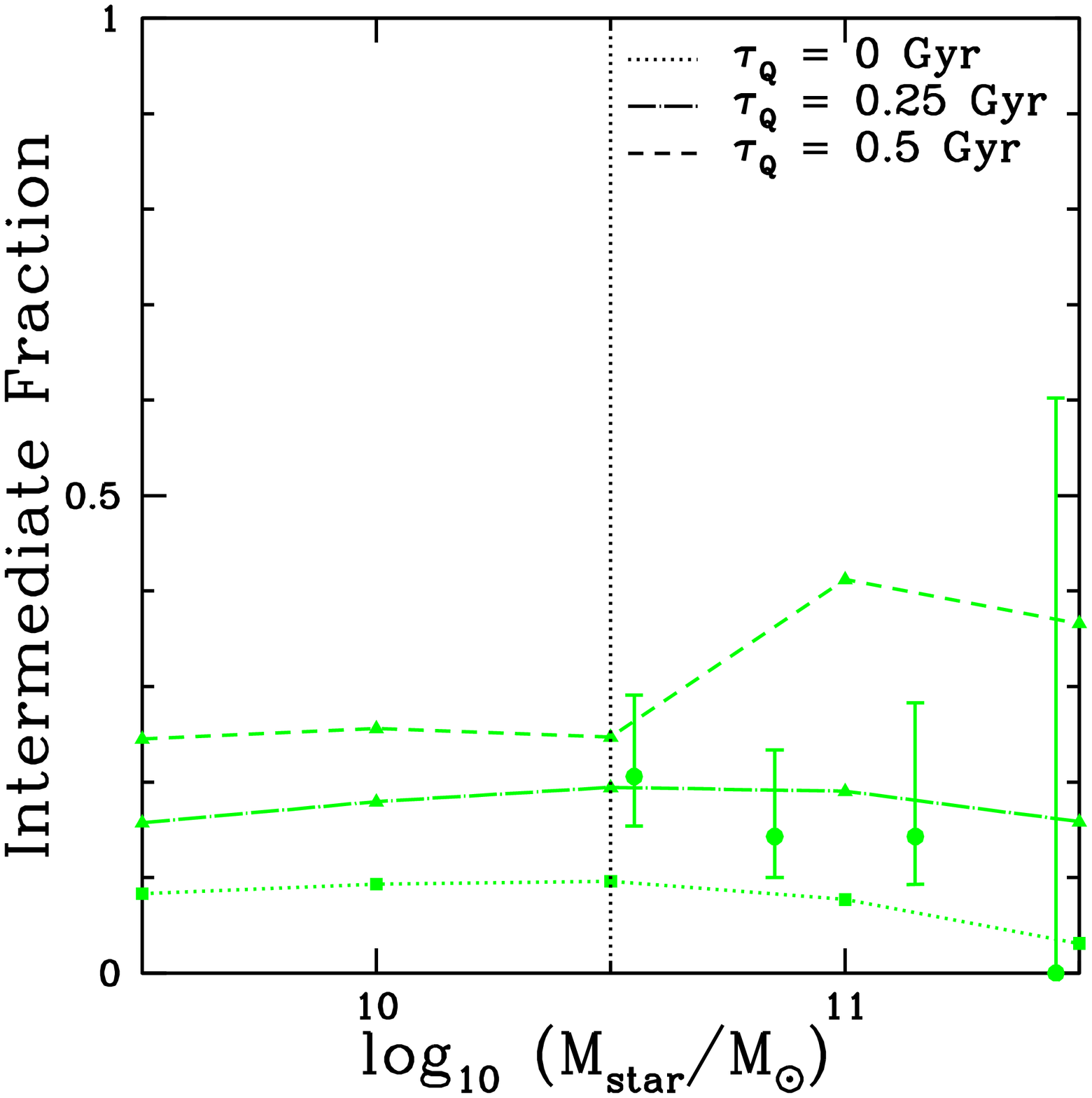}
	\epsfysize=5.5cm
	\epsfbox{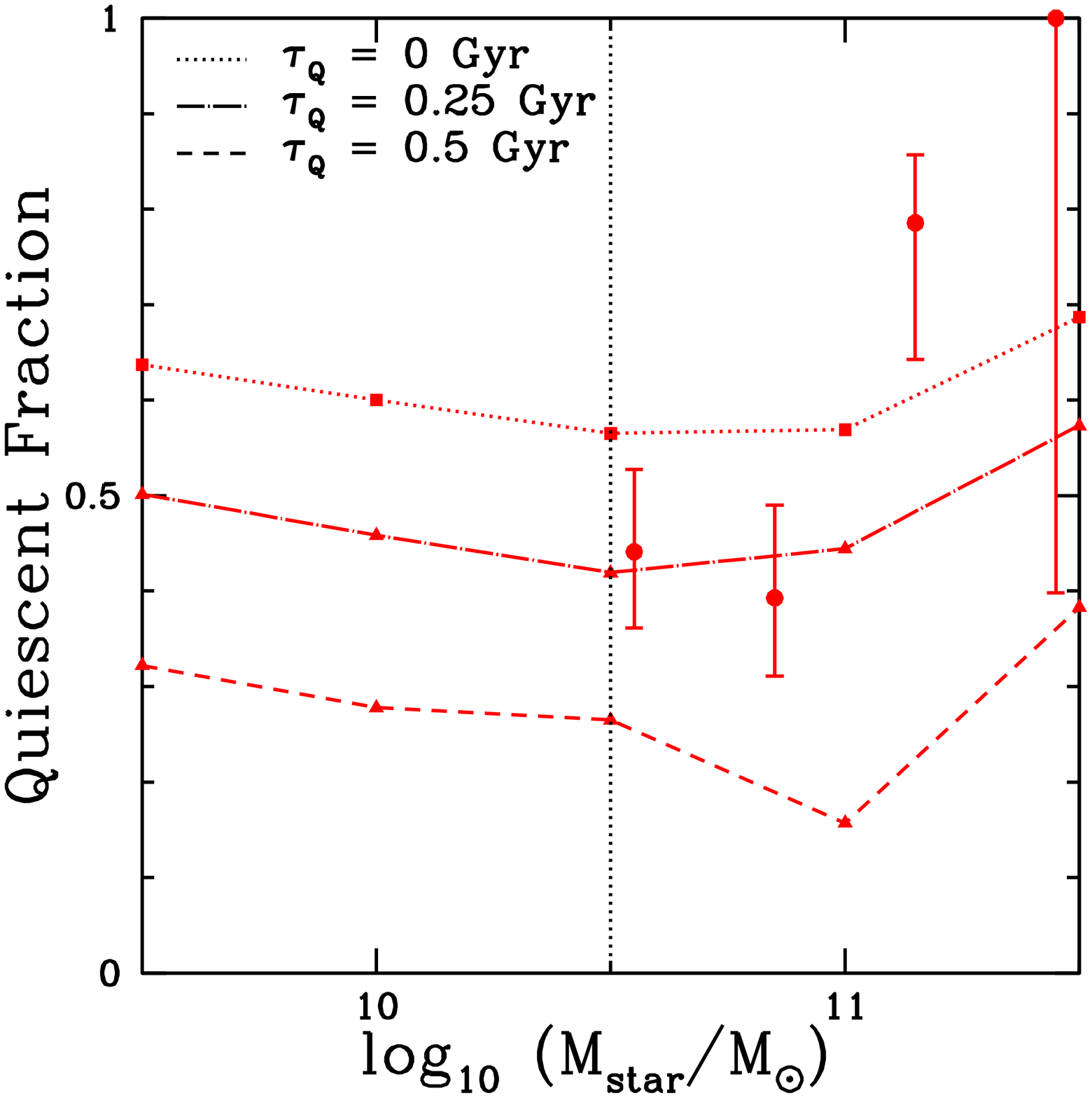}
	\caption{Satellite quenching model with varying $\tau_Q$ + {\bf a delay timescale that is dependent on the dynamical time} and its impact on the star-forming {\it (left)}, intermediate {\it (centre)}, and quiescent {\it (right)} galaxy fractions for the Behroozi models. The {\it filled triangles} indicate different mass bins, with {\it dotted line} connecting models with $\tau_Q=0$ Gyr, {\it thick dashed line} connecting models with $\tau_Q=0.25$ Gyr, and {\it thin dashed lines} connecting models with $\tau_Q=0.5$ Gyr. The observed group fractions from GEEC2 are presented as points with error bars. The black dotted line at $10^{10.5} M_\odot$ shows where the sample is reasonably complete, even for passive galaxies.}
	\label{fig-Model_DynDelay_Frac}
\end{figure*}
\begin{figure*}
	\leavevmode
	\epsfysize=8.5cm
	\epsfbox{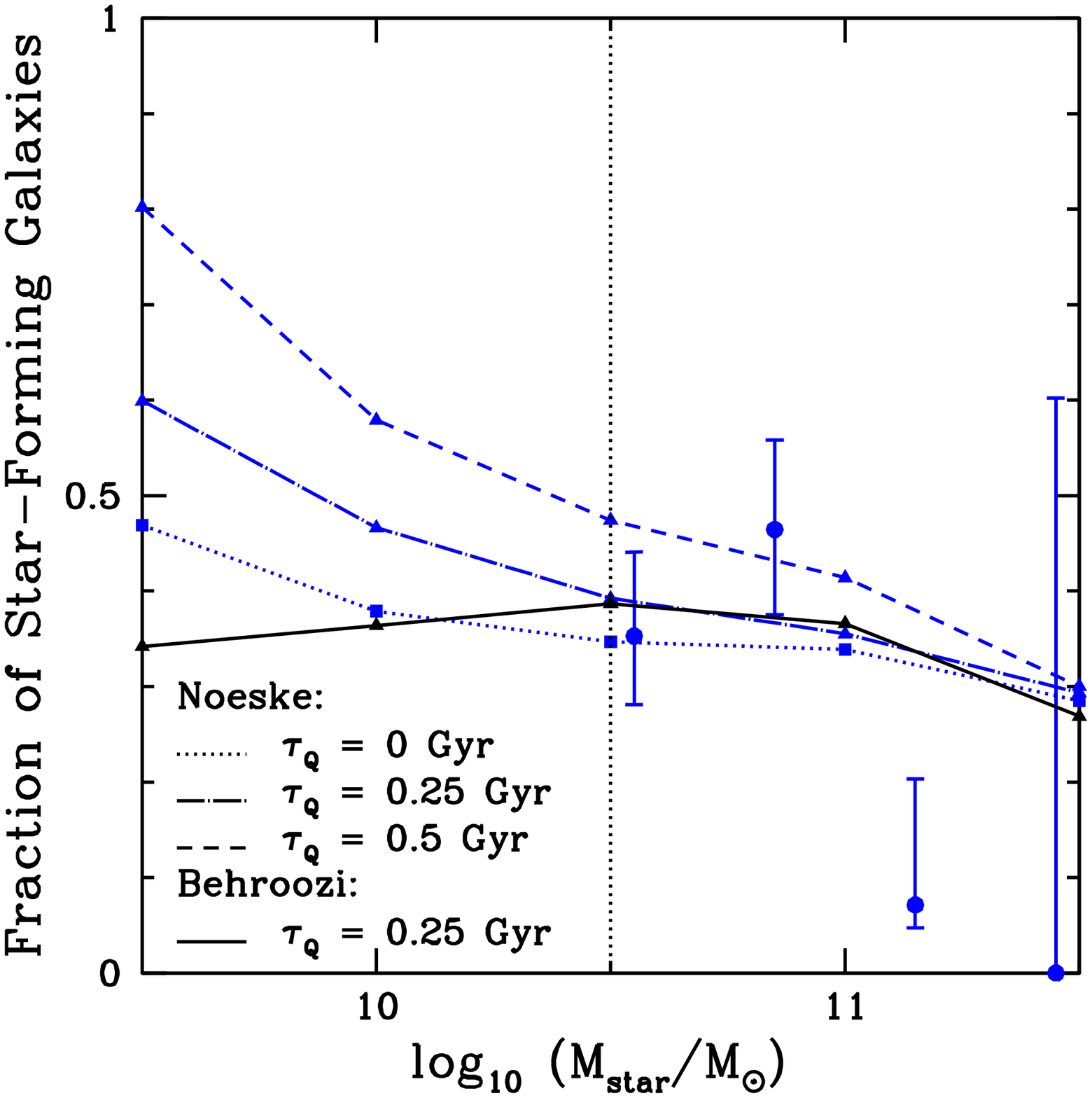}
	\epsfysize=8.5cm
	\epsfbox{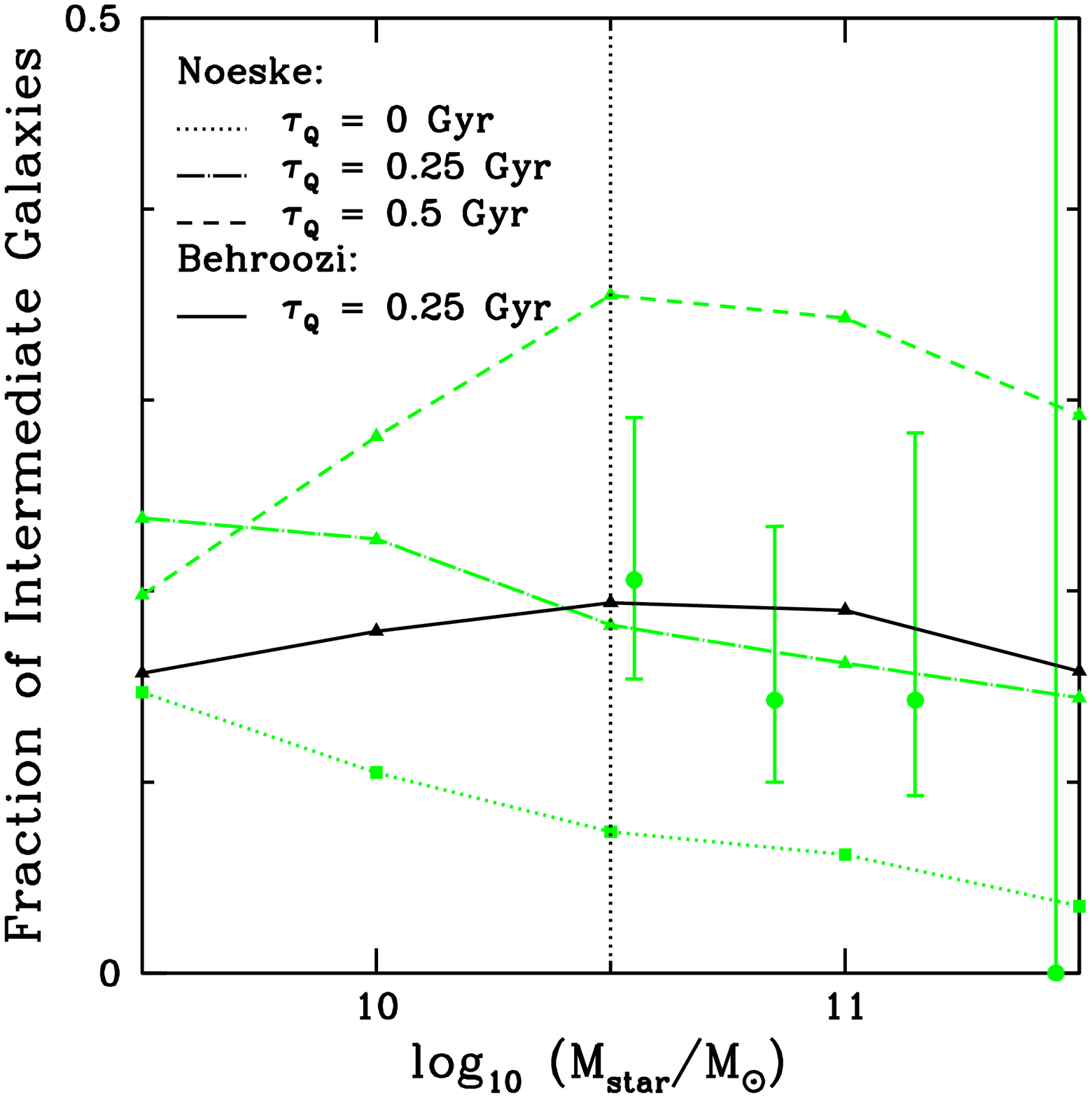}
	\caption{Comparison of the star-forming {\it (left)} and intermediate {\it (right)} fraction in Noeske's and Behroozi models for the satellite quenching model with varying $\tau_Q$ + a delay timescale that is dependent on the dynamical time. A {\it blue dotted line} connects Noeske models with $\tau_Q=0$ Gyr, a {\it blue thick dashed line} connects Noeske models with $\tau_Q=0.25$ Gyr, and a {\it blue thin dashed lines} connects Noeske models with $\tau_Q=0.5$ Gyr. The {\it black solid line} is the reference Behroozi model with $\tau_Q=0.25$ Gyr. The observed group fractions from GEEC2 are presented as points with error bars. The black dotted line at $10^{10.5} M_\odot$ shows where the sample is reasonably complete, even for passive galaxies..}
	\label{fig-compbluegreen}
\end{figure*}
\par
The next step is to introduce a delay, which means that a period of time must pass between the time of accretion of the satellite into the group and the beginning of the quenching process (i.e. where the exponential function takes over). This has been suggested, especially at $z=0$ to match the observed distribution of star formation rates \citep{2012MNRAS.424..232W, 2013MNRAS.432..336W} and the quiescent fraction \citep{2012MNRAS.423.1277D}.
\subsubsection{3 Gyr Delay}
\par
A 3 Gyr delay was added to the model, as suggested for the local universe by \citet{2012MNRAS.424..232W}. We maintained the previous values for the quenching timescale, $\tau_Q=0$ Gyr (immediate cut-off), $\tau_Q=0.25$ Gyr, and $\tau_Q=0.5$ Gyr. The result of this change can be seen in Figure~\ref{fig-Model_3Delay_Frac}.
\par
For all cases, not enough quiescent galaxies have been produced by the time of observation, since a large proportion of satellite galaxies would not have begun to feel the quenching process and therefore would stay on the star-forming (`blue') sequence. The formation timescale of these model galaxies at $z=7.95$ and the age of the universe at $z\sim1$ would likely prohibit such a long delay timescale.
\subsubsection{Delay Dependent on Dynamical Time}
\par
While the satellite quenching model with a $3$ Gyr delay may work well in the local universe, this value must be lower at $z\sim1$. Although a smaller value, such a $\sim1$ Gyr may work for $z\sim1$, we have chosen to model the case where this delay scales with the dynamical time, as suggested by \citet{2010ApJ...719...88T}. The delay timescale would then evolve with redshift as $(1+z_{\rm infall})^{-\frac{3}{2}}$, which can be individually calculated for each galaxy at their time of accretion. Once again, the three quenching timescale are kept, $\tau_Q=0.0$ Gyr (immediate cut-off), $\tau_Q=0.25$ Gyr, and $\tau_Q=0.5$ Gyr. The resulting colour fractions are presented in Figure~\ref{fig-Model_DynDelay_Frac}.
\par
The star-forming and quiescent fractions appear to be much better matched to the observations than the first two scenarios. A delay timescale evolving with the dynamical time could explain the star-forming and quiescent population, while with the correct value of $\tau_Q$, this model could also produce the correct fraction of intermediate galaxies. From the data, the $\tau_Q=0.25$ Gyr seem to be most likely, with the $\tau_Q=0.5$ Gyr or $\tau_Q=0.0$ Gyr models performing worse, as they tend to over-produce or under-produce intermediate (`green') galaxies respectively.
\par
Another point to note is that the star-forming and quiescent fractions in the Behroozi models do not appear to be strongly mass-dependent, as suggested by the GEEC2 results and other similar observations. This may be an indication that the different star formation histories for galaxies of varying stellar mass do not have a significant influence on the quiescent fraction, when compared to the quenching mechanism. One possible method of reconciling between the observations and the results from the modelling may be a mass-dependent $\tau_Q$ or t$_Q$, but this is beyond the scope of this paper. In addition, if higher mass galaxies are preferentially quenched, such as in Peng's model, then it could be another method of creating mass-dependence in the quiescent fraction.
\subsubsection{Comparison with Noeske's Models}
\par
The Noeske star formation histories provide similar results for the scenario with a delay dependent on the dynamical time, as shown in Figure~\ref{fig-compbluegreen} for the fraction of star-forming and intermediate galaxies. The similar results for the intermediate galaxies galaxies is to be expected, since the proportion is strongly dependent on the exponential quenching timescale ($\tau_Q$).
\par
The more interesting result comes from the fraction of star-forming galaxies. The mass dependence in the star-forming fraction with the Noeske models comes from the different formation time of satellite galaxies, which leads to differences in the steepness of the accretion curve, as seen previously in Figure~\ref{fig-infall}. Therefore, the stronger mass dependence in the fraction of quiescent and star-forming galaxies comes from the varying proportion of galaxies that had `felt' the effects of quenching. On the whole, the results seem to match the Behroozi models in the high mass end, with both sets fitting the GEEC2 data points. However, these two models do make different predictions in the low mass end, which could be tested with future deep observations.
\subsection{\hdelta Analysis}
\par
We consider the strength of the \hdelta feature for the different models in Figure~\ref{fig-Model_DynDelay_VZHD}, by performing the same weighted averages for the fraction of quiescent, intermediate and star-forming galaxies. We then take the observed values from the GEEC2 survey to compare to our results. To calculate the \hdelta strengths, we first create a stacked spectra for three categories and then calculate the equivalent width based on the bandpass method. The full outline can be found in \citet{2013MNRAS.431.1090M}.
\par
The quenching models produce \hdelta strengths higher than from the GEEC2 sample of intermediate (`green') galaxies. This is a roughly $2\sigma$ deviation to the $\tau_Q=0.25$ Gyr models). The relatively large uncertainties in this measurement are caused by our small sample of these `green' galaxies as well as the low signal-to-noise measurements of their spectra.
\par
Interestingly, the models also predict higher \hdelta strengths for the star-forming galaxies as well. This may be caused by the underestimation of the emission component in the GEEC2 sample, as noted in \citet{2013MNRAS.431.1090M}. We have attempted to compensate for the emission feature observed in the spectral line by fitting it with a Gaussian function and then adding this component back into the total \hdelta equivalent width. Note that this procedure was done for the star-forming galaxies only, as there was no emission feature observed in the stacked spectra for the quiescent and intermediate population.
\par
There is also a relatively small number of massive, star-forming galaxies, which means that the stacked spectrum may not be indicative of the sample mean. In addition, the field \hdelta strengths from Figure 16 of \citet{2013MNRAS.431.1090M} seems to be a better match for the model star-forming galaxies, especially the low-mass bin. This may indicate that the massive star-forming galaxies have already been affected by the environment or other sources of quenching, in a way that is not captured by this simple model.  However, given the low numbers of galaxies and the systematic uncertainties due to emission-filling and telluric contamination, we hesitate to over-interpret the low \hdelta of massive, star-forming group galaxies.
\par
Next, in order to reconcile between observations and the quenching models for intermediate galaxies, there are several ways to create galaxies with more moderate \hdelta strengths. In the context of the quenching model, this can be accomplished by having a longer $\tau_Q$, as seen in Figure~\ref{fig-Model_DynDelay_VZHD}. It is also possible to create more moderate \hdelta strengths by having a distribution for $\tau_Q$ instead of a single value.
\par
Another way of reducing the \hdelta strength of the intermediate population is if not all the intermediate galaxies are created through the quenching process. A method of creating intermediate galaxies is the process of mass-quenching, such as in the model of \citet{2010ApJ...721..193P}. These luminous galaxies will likely dominate the \hdelta signal observed. Some other methods of creating moderate \hdelta strength intermediate galaxies include the rejuvenation of early-type galaxies and quenched galaxies possessing residual star formation. These two specific model will be investigated through stellar synthesis models.
\subsubsection{Rejuvenation Model}
\begin{figure}
	\leavevmode
	\epsfysize=8.5cm
	\epsfbox{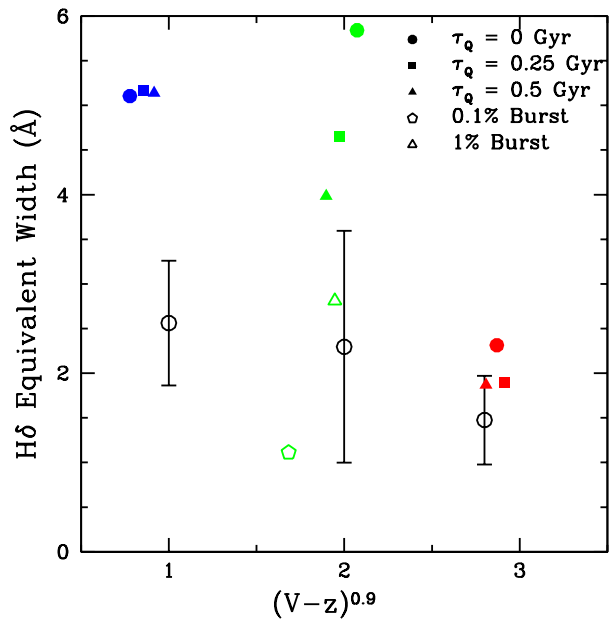}
	\caption{\hdelta equivalent widths vs $(V-z)^{0.9}$ colour for satellite quenching models with a varying $\tau_Q$ and a delay timescale that is dependent on the dynamical time. The results for models with a mass of $\log(M_*/M_\odot)=10.5$ are presented. The {\it filled circles} indicate the population of quiescent, intermediate and star-forming galaxies with $\tau_Q=0$ Gyr, the {\it filled squares} for models with $\tau_Q=0.25$ Gyr, and the {\it filled triangles} for models with $\tau_Q=0.5$ Gyr. The {\it open pentagon} shows the expected \hdelta value for the intermediate galaxies with a $0.1$ per cent burst, while the {\it open triangle} show the same for the $1$ per cent burst. The {\it open circles} with the error bars are the measured \hdelta widths from the GEEC2 survey (their locations on the x-axis is arbitary), which are taken from Figure 16 of \citet{2013MNRAS.431.1090M}. The value for the star-forming galaxies is from the high-mass sample ($\log(M_*/M_\odot)>10.3$).}
	\label{fig-Model_DynDelay_VZHD}
\end{figure}
\begin{figure*}
	\leavevmode
	\epsfysize=5.5cm
	\epsfbox{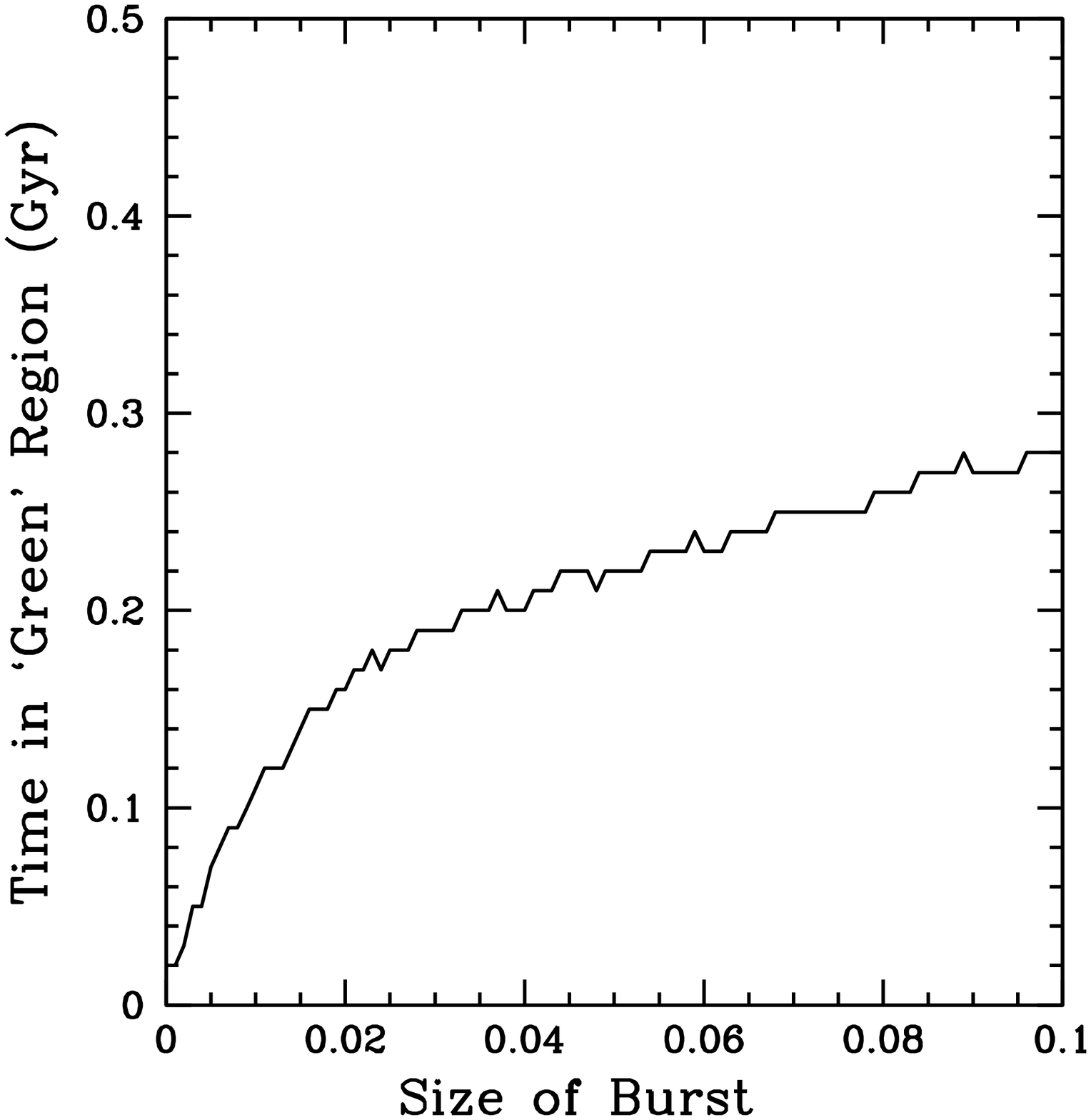}
	\epsfysize=5.5cm
	\epsfbox{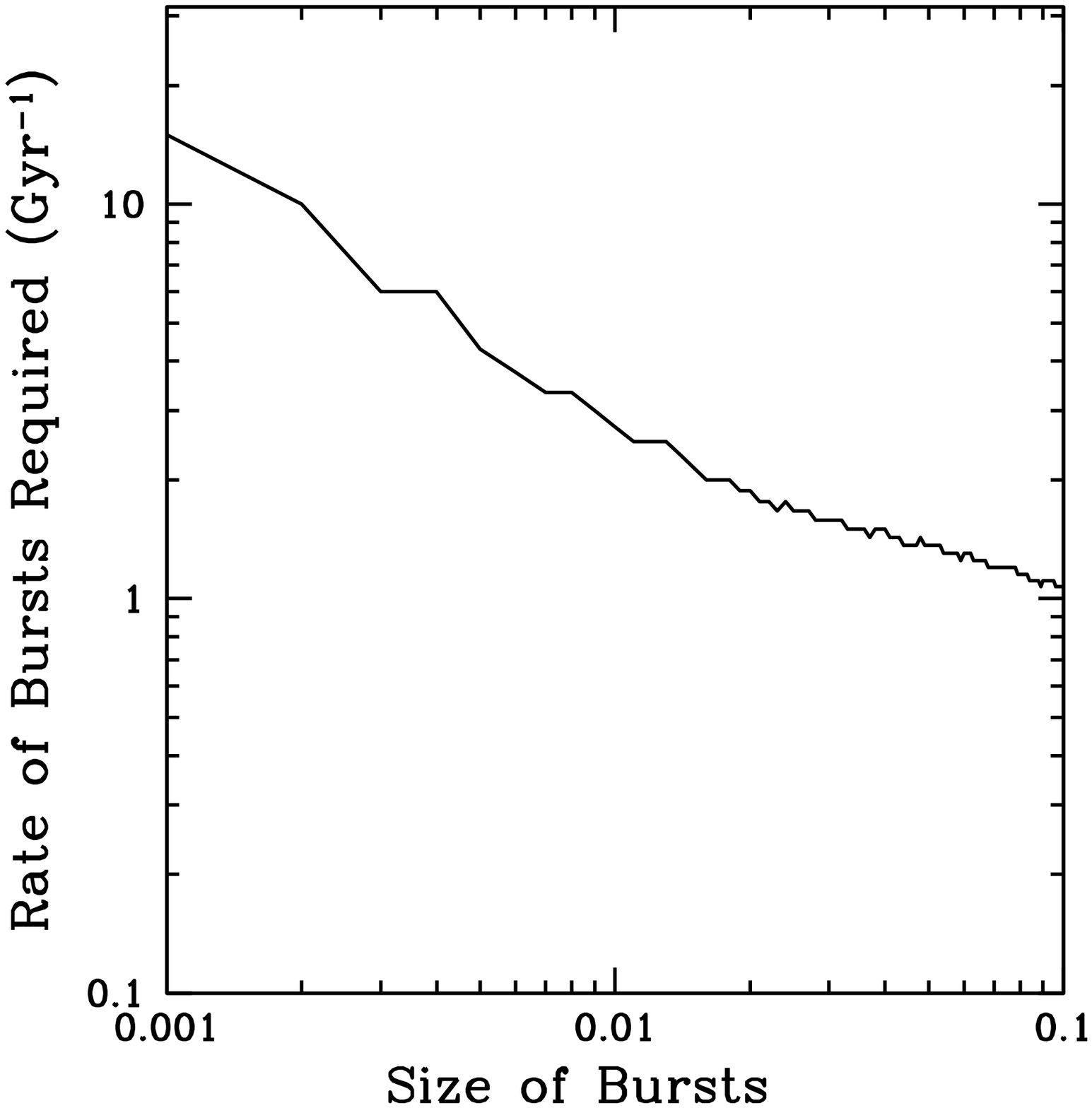}
	\epsfysize=5.5cm
	\epsfbox{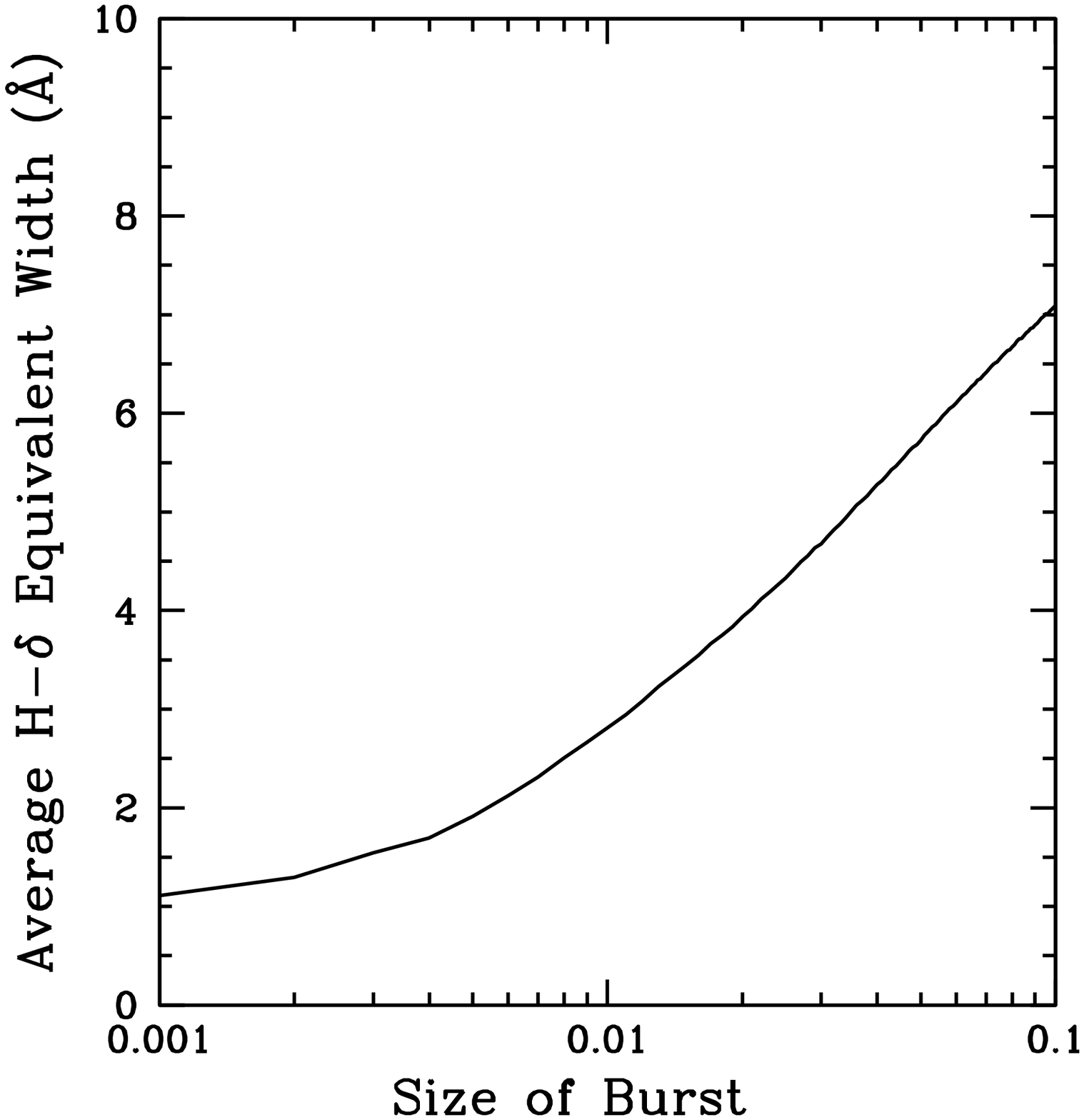}
	\caption{Summary of results from the rejuvenation model. {\it Left:} The length of time that the rejuvenated galaxy is observable in the `green' region, as a function of the size of the burst (as a fraction of the original mass of the host galaxy). {\it Centre:} The number of bursts required per Gyr to explain the proportional of intermediate galaxies in the GEEC2 survey, as a function of the size of the burst. {\it Right:} The average \hdelta strengths of these rejuvenated galaxies, as a function of the size of the burst.}
	\label{fig-Model_T02}
\end{figure*}
\par
To attempt to explain this potential discrepancy in the \hdelta strength, we test another hypothesis for the creation of these intermediate (`green') galaxies. We can assume that these intermediate galaxies are not formed through the process of quenching satellite galaxies, but rather from the rejuvenation of star formation in red, passive galaxies. For example, this can occur through wet minor mergers or through gas accretion \citep{2010ApJ...714L.290S, 2012ApJ...761...23F}.
\par
Since these intermediate galaxies are a subset of the quiescent population, we can use stellar synthesis models to constrain the size and frequencies of the required bursts of star formation, based on the observed properties of these intermediate galaxies and the expected timescale that such rejuvenated early-type galaxies spend inside the `green' region. 
\par
To create `passive' galaxies with the stellar synthesis models, we chose the simplest route. This can be done by starting with an initial burst of star formation in the beginning, and then zero star formation thereafter. Intermediate (`green') galaxies are formed by adding in a smaller burst of star formation 3 Gyrs later. As a result, these models will allow us to constrain the proportion of passive galaxies that would be required to undergo this process, as well as the required properties of these bursts.
\par
The results of the rejuvenation model can be seen in Figure~\ref{fig-Model_T02}. First, we find that the length of time in the `green' region is a smooth function of the burst size, which ranges from $\sim0.02$ Gyr for a $0.1$ per cent burst to $\sim0.3$ Gyr for a $10$ per cent burst. This can then be converted to an estimate of the number of bursts required to create the population of intermediate galaxies observed. For example, a $\sim1$ per cent burst would spend $\sim0.1$ Gyr in the `green' region. Since intermediate galaxies compose up to $20$ per cent of the overall galaxy population or up to $30$ per cent of the total combined quiescent + intermediate population in the GEEC2 sample, this would mean that these quiescent galaxies have to switch to the `green' phase $\sim3$ times per Gyr. We can compute that value for each individual model in Figure~\ref{fig-Model_T02}, where the number of bursts required ranges from $\sim10$ for a $\sim0.1$ per cent burst to only $\sim1$ for a $\sim10$ per cent burst.
\par
From Figure~\ref{fig-Model_T02}, we can see that this value ranges from $\sim1.6$\AA\ for a $\sim0.1$ per cent burst to $\sim6.5$\AA\ for a $\sim10$ per cent burst. For example, the $10$ per cent burst would make the galaxy extremely strong in the \hdelta absorption feature, far stronger than the range we would expect from our GEEC2 `green' group sample. Note that a $\sim1$ per cent burst would require roughly a maximum of 3 events per Gyr (assuming all the observed `green' galaxies are the rejuvenated variety), which may not be completely unrealistic at this redshift and mass ratio \citep{2008ApJ...681..232L}.
\par
Referring back to Figure~\ref{fig-Model_DynDelay_VZHD}, while the sample of intermediate galaxies is small, and the measurement of this weak line is generally of low signal to noise, it appears that the rejuvenation model (with a $1$ per cent burst) does predict a more reasonable range of \hdelta values. This is likely because the relatively small sizes of the burst compared to the stellar population, while enough to push a galaxy into the `green' window, is not enough to lead to a significant increase in the \hdelta strength.
\subsubsection{Quenching Models with Residual Star Formation}
\par
Another method of reconciling between the relative fraction of intermediate galaxies and their \hdelta strengths in the quenching model may be through these quenched galaxies having some residual star formation. For example, we can model this by introducing a `floor' or a lower limit to the star formation rate.
\begin{figure}
	\leavevmode
	\epsfysize=8.5cm
	\epsfbox{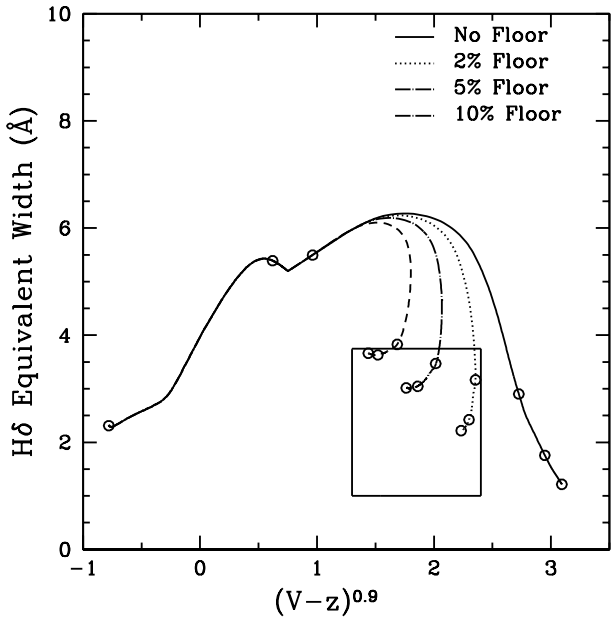}
	\caption{The evolutionary track of quenched galaxies with different levels of residual star formation ($0$ per cent, $2$ per cent, $5$ per cent, and $10$ per cent) in a plot of the \hdelta equivalent widths vs $(V-z)^{0.9}$ colour. The galaxies have a constant star formation rate for 2 Gyr, which is followed by an immediate quenching process. The {\it open circles} indicate time intervals of 1 Gyr. The {\it black box} shows the colour selection for the intermediate population from our models and the observed \hdelta equivalent width from the GEEC2 survey.}
	\label{fig-Model_T08Floor_VZHD}
\end{figure}
\par
We can illustrate the influence of a floor in star formation by the use of a simple model, where a galaxy starts with an initial 2 Gyr phase of constant star formation, followed by an instantaneous cut-off and letting the object evolve for another 3 Gyr. We can then vary the star formation rate floor and determine the impact on the colour and \hdelta strength on the resulting objects. The evolutionary tracks are presented in Figure~\ref{fig-Model_T08Floor_VZHD}, with a plot of \hdelta equivalent widths vs $(V-z)^{0.9}$ colour for galaxies with different floors.
\par
In Figure~\ref{fig-Model_T08Floor_VZHD}, we see that it is possible to create intermediate galaxies with a good choice of the star formation rate floor. With a residual star formation rate of $\sim2-10$ per cent, we can create galaxies that stay in the intermediate region for $\gtrsim2$ Gyr. Furthermore, these galaxies possess relatively low \hdelta strengths, compared to the intermediate galaxies created by the original quenching models.
\par
In the context of the original quenching model, it may be possible to combine the instantaneous cut-off scenario with a star formation rate floor. Assuming that the timescale is insufficient to create any `blue' galaxies through this process, then around $\sim10$ per cent of the quiescent galaxies observed at $z\sim1$ would have to possess this floor to create a sufficient number of intermediate galaxies. As these new objects would comprise up to half of the `new' intermediate population, this may be sufficient to reduce the average \hdelta strength by a sufficient amount to be consistent with the GEEC2 observations.
%
%
\section{Discussion}
%
\par
With the combination of observations and the results presented in the previous section, we can begin to determine which models are favoured and which models can be safely excluded. From the perspective of the GEEC2 survey, the number of galaxies in our sample, and the mass range where we are complete, the choice of the Behroozi and Noeske star formation histories have led to the same conclusion.
\par
For the quenching models we have chosen, a combination of a delay timescale that scales with the dynamical time and a quenched function with an exponential timescale ($\tau_Q$) of $\sim0.25$ Gyr seems to best match the observed fractions. Models with a lower value of the delay timescale, such as the no delay models in this paper, would require a much larger $\tau_Q$ to reduce the number of quiescent galaxies. However, $\tau_Q$ is strongly constrained by the proportion of intermediate galaxies observed. On the other hand, a very long delay timescale, such as a constant 3 Gyr from the local universe, may be hard to reconcile with observed fractions of quiescent galaxies at $z\sim1$.
\par
The low observed \hdelta presents a puzzle, as it seem to preclude such a rapid quenching timescale. One solution would be that only some of the intermediate galaxies are actually in transition and the others could be the result of mass quenching, rejuvenation, or other processes. For example, our quenching model  also does not include the influence of any mass-quenching component, which would affect the star-forming/quiescent fractions for the more massive galaxies in our sample. However, our conclusions from the quenching models remain valid based on the two lowest mass bins.
\par
Furthermore, the \hdelta signal will also be affected by the mass-quenched population, which may contain galaxies with lower \hdelta strengths compared to the environmentally quenched population. In that case, the stacked \hdelta signal from the GEEC2 sample, weighted by light, could be reflecting the influence of this mass-quenched population. This leads to another possible solution to the discrepancy in the \hdelta strength.
\par
If these alternative sources of intermediate galaxies are present, then the amount of `transition' or environmentally quenched intermediate galaxies would have to be even lower. Therefore, the exponential timescale ($\tau_Q$) can be constrained to be $<0.25$ Gyr. Another possibility is the presence of a small number of quenched galaxies with residual star-formation, which can produce long-lasting intermediate galaxies with moderate \hdelta strength.
\par
A caveat and possible extension to this method is that we have assumed all galaxies at fixed stellar mass and redshift have the same star formation rates. There are of course stochastic variations in the star formation rate of galaxies about the average, which may affect the resulting fractions of star-forming, intermediate, and quiescent galaxies. However, this effect should be small given the tightness of the fit to the star-forming sequence from the GEEC2 data. Another caveat is the effect of projection and interlopers on the quiescent fraction in our GEEC2 sample. Other groups have found the purity of satellites at a slightly lower redshift to be $\sim80$ per cent \citep{2013arXiv1307.4402K}. Therefore, it is probable that the quiescent fraction is an underestimate of the true value and a shorter t$_Q$ could be tolerated.
\par
For future observations, there are two main avenues to pursue. First, we will need to increase the signal-to-noise ratio for \hdelta measurements and to reduce the error bars for the measurements of the fraction of quiescent, intermediate and star-forming galaxies for our group galaxies. This can be accomplished by simply observing more groups and identifying more satellite galaxies. Having a larger sample would allow us to further constrain the parameters of any simple quenching or rejuvenation model.
\par
Second, we will need to go deeper in our surveys at $z\sim1$, in order to find lower mass satellite galaxies. This is especially true for the quiescent galaxies and the intermediate (`green') population, as the GEEC2 survey is only complete at a much higher stellar mass limit compared to actively star-forming galaxies \citep{2013MNRAS.431.1090M}. A push to lower mass satellites will allow us to differentiate between the mass-dependent quenching models and the mass-independent rejuvenation models. In addition, Noeske and Behroozi star formation histories predict different trends of the fraction of quiescent and star-forming galaxies with mass. If we can determine the fractions of quiescent, intermediate and star-forming satellites galaxies with stellar masses less than $10^{10.5}M_\odot$, we can begin to verify the predictions that our models have made.
%
%
\section{Conclusions}
\par
We have presented new analysis of the data from the GEEC2 spectroscopic survey of group galaxies at $z\sim1$, including a discussion of their star formation rates as well as an overview of the properties of the observed group galaxies. For the quenching models, we used the \citet{2012arXiv1207.6105B} and \citet{2007ApJ...660L..47N} star formation histories to model star-forming galaxies. We tested the use of different quenching timescales ($\tau_Q$), and the addition of delays (t$_Q$), including a constant 3 Gyr delay and a delay dependent on the dynamical time. We conclude the following:
\begin{itemize}
	\item The star-forming group galaxies seem to match the star forming sequence for field galaxies at $z=0.9$ of \citet{2012ApJ...754L..29W}, especially in the region where the GEEC2 sample is complete ($>10^{10.5} M_\odot$).
	\item For the FUV measurements, if we take the limit of $4\times$ sSFR($z$) threshold from \citet{2011ApJ...739L..40R} to be the threshold for galaxies showing a significant excess in star formation, then only $4.4\substack{+3 \\ -1}$ per cent fit this criteria (4 out of 90), as compared to $5.1\pm1$ per cent (44 out of 871) in the field. From the \oii star formation rate measurements, only $6.7\substack{+4 \\ -2}$ per cent of group galaxies fit this criteria (6 out of 90). This is in agreement with results from our previous paper, where we observed no significant difference in the star formation rates of star-forming group and field galaxies. This result also gives validity to our assumption that quenching is the primary factor in the transition between star-forming and quiescent states.
	\item For the quenching scenario, the fraction of intermediate (`green') galaxies is strongly dependent on the exponential quenching timescale ($\tau_Q$). The star-forming/intermediate/quiescent fractions rule out the no delay scenario, which would require a long $\tau_Q$ that over-produces immediate-colour galaxies. They also rule out the 3 Gyr delay model, which does not produce a sufficient amount of quiescent galaxies. The observed fractions are best matched with a dynamical delay time and $\tau_Q=0.25$ Gyr.
	\item The best-fitting quenching model predicts the intermediate (`green') galaxies \hdelta strength to be higher than the observed \hdelta strengths ($\sim2\sigma$ deviation), but we do caution that there is small number of these `green' galaxies in the GEEC2 sample and the low signal to noise ratios of their spectra. However, this result does suggest that a significant portion of the observed intermediate galaxies may have originated from other sources.
	\item The combination of environmental quenching and other processes may be at work in the GEEC2 sample, including the presence of a mass-quenched population. If the population of intermediate galaxies had originated from a combination of these sources, then we can constrain $\tau_Q\lesssim0.25$ Gyr.
	\item Two other scenarios were explored with stellar synthesis models, including rejuvenated early-type galaxies and quenched galaxies with residual star formation. We note that a large number of small bursts or a smaller number of large bursts can both explain the population of intermediate (`green') galaxies, though the observed \hdelta strength is best matched with a burst size of $\sim1$ per cent and a burst rate of $\sim3$ times per Gyr. From the residual star formation scenario, a star formation rate floor of $\sim2-10$ per cent would allow quenched galaxies to maintain their position in the intermediate region for $\gtrsim2$ Gyr, with moderate \hdelta strengths that are better matched with observations.
\end{itemize}
\par
The models in this paper, although simple, can be used as a starting point to a better understanding of the star-formation histories of these group galaxies. From the results, we know that environmental quenching must act quickly once it has begun, in order to explain the fraction of intermediate galaxies observed. The models also suggest that any delay at $z\sim1$ must be shorter than at $z=0$, which indicates that this process may be related to the dynamical time or to the orbit of galaxies in the haloes.
%
%
\section{Acknowledgments}
\par
We are grateful to the COSMOS and zCOSMOS teams for making their excellent data products publicly available. This research is supported by NSERC Discovery grants to MLB and LCP. MLB acknowledges grants from NOVA and NWO, which supported his sabbatical visit at Leiden Observatory, where this work was completed. RGB is supported by STFC grant ST/I001573/1. The data used in this work can be downloaded from the Gemini data archive.
%
%
\bibliographystyle{mn2e}
\bibliography{paper2}
%
%
\end{document}